\documentclass[times,final,nopreprintline]{elsarticle}

\usepackage{framed,multirow}
\usepackage[a4paper,top=3cm,bottom=2cm,left=2.5cm,right=2.5cm,marginparwidth=1.75cm]{geometry}

\usepackage{amssymb}
\usepackage{latexsym}
\usepackage{amsmath}
\usepackage{physics}
\usepackage{amsfonts}
\usepackage{bm}
\usepackage{subcaption}
\usepackage{url}
\usepackage{xcolor}
\definecolor{newcolor}{rgb}{.8,.349,.1}

\usepackage{algorithm}
\usepackage{algpseudocode}
\usepackage{hyperref}

\newcommand{\appref}[1]{%
  \hyperref[#1]{Appendix}%
}

\begin{document}

\begin{frontmatter}

\title{Reconstructing the Magnetic Field in an Arbitrary Domain via Data-driven Bayesian Methods and Numerical Simulations}

\author[1]{Georgios~E. Pavlou \corref{cor1}}
\cortext[cor1]{Corresponding author: e-mail: gepavlou@iesl.forth.gr}
\author[2,3]{Vasiliki Pavlidou}  
\author[1,4,5]{Vagelis Harmandaris}

\address[1]{Institute of Applied and Computational Mathematics, Foundation for Research and Technology- Hellas, Heraklion, GR-71110, Greece}
\address[2]{Institute of Astrophysics, Foundation for Research and Technology- Hellas, Vasilika Vouton, Heraklion, GR-71110, Greece}
\address[3]{Department of Physics, University of Crete, Heraklion, GR-70013, Greece}
\address[4]{Department of Mathematics and Applied Mathematics, University of Crete, Heraklion, GR-71409, Greece}
\address[5]{Computation-Based Science and Technology Research Center, The Cyprus Institute, Nicosia, 2121, Cyprus}

\begin{abstract}
Inverse problems are prevalent in numerous scientific and engineering disciplines, where the objective is to determine unknown parameters within a physical system using indirect measurements or observations. The inherent challenge lies in deducing the most probable parameter values that align with the collected data. This study introduces an algorithm for reconstructing parameters by addressing an inverse problem formulated through differential equations underpinned by uncertain boundary conditions or variant parameters. We adopt a Bayesian approach for parameter inference, delineating the establishment of prior, likelihood, and posterior distributions, and the subsequent resolution of the maximum a posteriori problem via numerical optimization techniques. The proposed algorithm is applied to the task of magnetic field reconstruction within a conical domain, demonstrating precise recovery of the true parameter values.
\end{abstract}


\end{frontmatter}

\section{Introduction}
\label{intro}

Inverse problems are encountered across a vast spectrum of scientific and engineering disciplines, from~astrophysics to medical imaging, and~their reach extends to geophysics and non-destructive testing, among~others. Inverse problems are framed by the challenge of deducing unknown system properties or parameters from indirect, often noisy, and~uncertain experimental or observational data~\cite{Aster2013, Bissantz2008}. Moreover, inverse problems are sometimes connected to reconstruction problems, where one wants to calculate a physical quantity in a whole domain, which is only partially known from experiments. Therefore, the~intrinsic ill-posedness of inverse problems complicates the search for accurate and robust solutions and introduces complexities in numerical reconstruction efforts, where the goal is to compute physical values throughout a domain based on partial data~\cite{Arridge2019, Kabanikhin2020}.

The principal objectives of this work are to develop a versatile, data-driven Bayesian algorithm for solving inverse problems with applications in reconstructing physical quantities, which are calculated from a forward problem, usually as a solution of a differential equation, from~sparse data. By~integrating Bayesian inference with clustering techniques, we can robustly infer boundary conditions and provide detailed reconstructions of the observed physical quantity in the given domain. Our results are supported by figures that elucidate the performance of our algorithm in various scenarios regarding the reconstruction of the magnetic field (MF) in a specific, arbitrarily defined~domain.

Statistical methods for addressing inverse problems have proliferated, with~frequentist and Bayesian approaches providing distinctive strategies for parameter estimation and uncertainty analysis. More particularly, Bayesian methods~\cite{Stuart2010} offer a structured probability-based framework that synthesizes prior information with observed data to articulate the posterior distribution of a parameter, which inherently captures the uncertainty of parameter inference and helps navigate the multidimensional parameter space. The~increasing applications of inverse problems amplify the call for novel methods. Challenges include managing high-dimensional parameters, computational burdens, and~the need for sophisticated algorithms designed to process large amounts of data efficiently~\cite{Cui2016}.

Reconstruction techniques are a fundamental aspect of computational science aimed at creating comprehensive models from incomplete or indirect data. These techniques are essential in transforming observational measurements into detailed representations of physical systems. Methods that resolve inverse problems are critically important, as~they provide a means to infer the full spectrum of system parameters, some of which may not be directly observable. Our objective is to accurately reconstruct the spatial distribution of physical quantities. For~example, in~geophysics~\cite{Snieder1999} and astrophysics~\cite{Vogt2005}, a~wide range of methodologies are applied to reconstruct the MF. Techniques such as boundary element methods~\cite{Liebsch2022}, Taylor series expansions~\cite{Broeren2021}, polynomial reconstructions~\cite{Denton2022}, the~pulsed wire method~\cite{Baader2022}, full vector tomography~\cite{Donnelly2018}, and~approaches derived from information field theory~\cite{Tsouros2023, Tsouros2024} have advanced MF reconstruction.
Moreover, recent advances in physics-informed machine learning have shown great promise in solving inverse problems by leveraging physical laws as prior knowledge within the learning process. These methods combine the strengths of data-driven approaches with the robustness of physics-based models, leading to improved interpretability and generalization of solutions~\cite{Raissi2019, Karniadakis2021}. Particularly noteworthy is their application in multi-scale modeling of molecular systems, where they facilitate the reintroduction of atomic detail in coarse-grained models. For~instance, a~recent work utilizes a physics-informed deep learning framework to accurately re-introduce atomic detail in coarse-grained configurations of multiple poly(lactic acid) stereoisomers, showcasing the potent utility of these methods in complex molecular reconstructions~\cite{Christofi2024}.

Despite these developments, there remains a need for a more versatile and generalized methodology that can be broadly applied across different domains. To~this end, sophisticated approaches that incorporate Bayesian inference algorithms stand out, as~they facilitate the assimilation of prior knowledge with observational data, yielding a robust framework for comprehensive reconstructions, regardless of whether we are interested in the MF or any other physical~quantity.

Building on this need, our paper presents a new data-driven Bayesian algorithm that is especially suited for inverse problems. This algorithm, which focuses on the particular challenge of identifying boundary condition parameters from diverse data sources, be it experimental, observational, or~synthetic, incorporates a probabilistic model that incorporates existing parameter knowledge and constructs a likelihood function to assess the model's consistency with observed data. By~utilizing Bayes' theorem, we orchestrate an amalgamation of prior knowledge and observational data, culminating in a well-founded probabilistic estimation of the parameters in question. The~reconstruction aspect is not merely a byproduct but~a targeted objective of our algorithm, aiming to provide a detailed and accurate portrayal of physical values across any given domain rather than only inferring internal parameters of partial differential~equations.

The utility of our algorithm extends beyond magnetic fields and serves as a robust tool for reconstruction in a range of scientific disciplines. For~example, in~medical imaging, particularly magnetic resonance imaging, reconstruction techniques often rely on solving inverse problems to create detailed images from raw data~\cite{Liang2020}. In~electrical engineering, inverse problems are integral to signal processing tasks, such as signal deconvolution, where a signal is reconstructed from incomplete frequency data~\cite{Tobar2023}. Geophysics uses inverse problems in seismic imaging to deduce the internal structure of the Earth from surface seismic wave measurements~\cite{Snieder1999}. In~astrophysics, inverse problems are crucial for tasks such as inferring the mass distribution of galaxy clusters from gravitational lensing data~\cite{Newbury2002}. Lastly, quantum tomography in quantum mechanics involves reconstructing quantum states or processes from measurement data, a~process that is typically framed as an inverse problem~\cite{Cao2022}.
\textcolor{black}{We consider this work as a first step towards reconstructing the galactic MF. The~algorithm and the applications presented in this work will form the basis for a realistic calculation where actual observational data will be used instead of artificial ones.}
The remainder of the paper is organized as follows. In~Section~\ref{math}, we present the mathematical formulation of the inverse problem and describe our algorithm in detail while also providing information on its various aspects and practical usage. In~Section~\ref{gmf}, we present how the forward problem is solved in our application. More specifically, we show how one can use the finite element method to calculate the MF in a given domain. In~Section~\ref{inv_gmf}, we apply our algorithm to the problem of reconstructing the MF in a domain. We demonstrate its effectiveness in synthetic data, showcasing its ability to accurately recover the true parameter values. We discuss the cases of multiple priors in which clustering of the data is necessary. Finally, Section~\ref{con} offers concluding remarks and discusses future perspectives. Furthermore, in the Appendix, we provide additional details on the formulation of the problem and the optimization~algorithm.

\section{Mathematical Formulation and Presentation of the~Algorithm}
\label{math}
\unskip

\subsection{Mathematical Framework for the Inverse~Problem}
Consider a physical system with known parameters $\mathbf{x}$ (which can, for~example, be the points on the grid of the physical domain $\Omega$ or other known quantities relevant to the problem) and unknown parameters $\boldsymbol{\theta}$ that can be measured by some indirect observation or experiment that produces the data set $\mathbf{y}$. The~relationship between the parameters $\mathbf{x}$, $\boldsymbol{\theta}$ and the observed data $\mathbf{y}$ can be described by a forward model $f(\mathbf{x};\boldsymbol{\theta} )$, which maps the parameters to the expected measurements. In~many cases, the~forward model involves solving a partial differential equation subject to certain boundary or initial conditions~\cite{Stuart2010}, where the unknown parameters, in~our case, can refer to any subdomain of $\Omega$ (for example, they may help us determine the boundary conditions) or to parameters that affect the initial conditions.
Mathematically, we can formulate the forward problem as follows~\cite{Kaipio2005}:
\begin{equation}
\label{first}
\mathbf{y} = f(\mathbf{x} ; \boldsymbol{\theta}) + \boldsymbol{\epsilon},
\end{equation}
where $\boldsymbol{\epsilon}$ is a noise term that accounts for measurement errors or~uncertainties. 

Suppose that from the observed data $\mathbf{y}$, we want to infer information about the statistical properties of the parameters $\boldsymbol{\theta}$ that determine the boundary conditions. This defines an inverse problem.
To be more concrete, consider a set of $n_y$, possibly sparse, measurements $\mathbf{y}$ randomly distributed in the known physical domain $\Omega$. In~many real-world scenarios, the~physical system and the observed data involve numerous variables and measurements that are of high dimension, while the domain geometry could also be large. Our algorithm is designed to handle high-dimensional problems, where, for~example, known parameters $\mathbf{x}$, $\mathbf{y}$ and~unknown parameters $\boldsymbol{\theta}$ can potentially have large dimensions. Furthermore, data $\mathbf{y}$ can be sparsely distributed in a large domain $\Omega$. Finally, the~parameters $\boldsymbol{\theta}$ can appear anywhere in the theoretical model; for example, they may appear in the differential equations that define the forward model, in~the boundary conditions, etc. Therefore, the~main challenge is to calculate the statistical properties of the parameters $\boldsymbol{\theta}$ using Bayesian~inference. 

In the context of this work, we assume that the physical system is described by a set of differential equations and that the unknown parameters $\boldsymbol{\theta}$ (treated as random variables) appear in the boundary conditions of the system. We focus on inferring the statistical properties of $\boldsymbol{\theta}$ rather than the boundary conditions themselves. To~improve the robustness and precision of our estimates, we repeat the optimization procedure multiple times with different initial guesses for $\boldsymbol{\theta}$ and~average the results. This approach helps to account for potential local optima or non-convexity of the posterior function and~allows us to explore the uncertainty and variability of the estimated parameters. Compared to previous work, which focused mainly on inverse problems with uncertain initial conditions or internal parameters of partial differential equations~\cite{Stuart2010, Hu2009, Yan2019, Padmanabha2021, Huang2022}, our approach addresses the relatively less explored area of inverse problems with uncertain boundary conditions. Therefore, our approach is distinct from previous work on inverse problems, as~we are specifically interested in the statistical properties of the parameters that affect the boundary~conditions.

\subsection{Bayesian Inverse~Problems}
Bayesian inference is a statistical framework that allows us to update our prior beliefs about the unknown parameters $\boldsymbol{\theta} =\left( \theta _1, \ldots,\theta _{n_{\theta}} \right) $, where $n_{\theta}$ is the number of parameters based on observed data $\mathbf{y}$ \cite{Stuart2010, Ghosh2006}. In~this context, we can use Bayesian inference to infer the statistical properties of unknown parameters that determine the boundary conditions of the physical system.
We start by specifying a prior probability distribution for the parameters $\boldsymbol{\theta}$, denoted as $p(\boldsymbol{\theta})$. The~choice of a prior distribution can incorporate our prior knowledge, beliefs, or~assumptions about the values of the unknown parameters. For~instance, we might choose a uniform if we have no prior information about the parameters or a normal distribution if we believe that they are centered around some central value. If~we have more than one parameter, and~assuming uncorrelated components of the parameter vector, the~total prior is the product of each one:
\begin{equation}
\label{prior1}
p(\boldsymbol{\theta} )=\prod_{i=1}^{n_{\theta}}{p(\theta _i)}.
\end{equation}

Given the observed data $\mathbf{y}$, the~posterior distribution over the parameters $\boldsymbol{\theta}$ is given by the Bayes' theorem:
\begin{equation}
\label{Bayes}
p(\boldsymbol{\theta} | \mathbf{y}) = \frac{p(\mathbf{y} | \boldsymbol{\theta}) p(\boldsymbol{\theta})}{p(\mathbf{y})},
\end{equation}
where $p(\mathbf{y} | \boldsymbol{\theta})$ is the likelihood function that describes the probability of observing the data $\mathbf{y}$ given the parameters $\boldsymbol{\theta}$, and~$p(\mathbf{y})$ is a normalization constant. The~likelihood function is typically obtained from the forward model $f(\mathbf{x};\boldsymbol{\theta})$ and the noise term $\epsilon$, assuming that observations are iid (independently and identically distributed), as follows:
\begin{equation}
\label{likelihood}
p(\mathbf{y} | \boldsymbol{\theta}) = \prod_{i=1}^{n_y} \frac{1}{\sqrt{2\pi\sigma^2}}\exp\left(-\frac{(y_i - f(\mathbf{x};\boldsymbol{\theta}))^2}{2\sigma^2}\right),
\end{equation}
where $n_m$ is the number of measurements, $\sigma$ is the standard deviation of the noise term, and~$y_i$ is the $i$th measurement in the dataset.

The posterior distribution $p(\boldsymbol{\theta} | \mathbf{y})$ provides a measure of the uncertainty or confidence we have about the values of the unknown parameters $\boldsymbol{\theta}$ after observing the data $\mathbf{y}$. We can compute various statistics of interest from the posterior distribution, such as mean, variance, or~credible intervals. In~the context of inverse problems, the~maximum likelihood estimate (MLE) is an alternative approach where the goal is to find the parameter values that maximize the likelihood function. MLE focuses solely on the likelihood, disregarding any prior information about the parameters. However, in~our Bayesian framework, we employ the maximum a posteriori (MAP) method, which extends the MLE concept by incorporating prior information. MAP maximizes posterior probability, offering a more comprehensive inference method that takes into account both the data (through the likelihood) and our prior knowledge (through the prior distribution). Specifically, our objective is to find the value of $\boldsymbol{\theta}$ that maximizes the posterior function:
\begin{equation}
\boldsymbol{\theta} _{\mathrm{MAP}}=\mathrm{arg}\max_{\boldsymbol{\theta}} (p\left( \boldsymbol{\theta} |\mathbf{y} \right) ).
\end{equation}

In practice, it is usually more efficient to maximize the logarithm of the posterior function as follows:
\begin{equation}
\label{mle}
\boldsymbol{\theta} _{\mathrm{MAP}}=\mathrm{arg}\max_{\boldsymbol{\theta}} (\ln p\left( \boldsymbol{\theta} |\mathbf{y} \right) )=\mathrm{arg}\min_{\boldsymbol{\theta}} (-\ln p\left( \boldsymbol{\theta} |\mathbf{y} \right) ),
\end{equation}
and $\boldsymbol{\theta}$ is calculated with proper numerical optimization techniques (more details will be given in the next Section) without having to evaluate the normalization constant $p(\mathbf{y})$.

We should note that alternatively to maximum a posteriori methods, Bayesian inference can be performed using Markov chain Monte Carlo (MCMC) methods~\cite{Brooks2011}.
MCMC algorithms draw samples from the posterior distribution $p(\boldsymbol{\theta} | \mathbf{y})$ without having to evaluate the normalization constant $p(\mathbf{y})$. The~samples can be used to estimate statistics of interest or to perform model selection and~comparison. 

In summary, Bayesian inference provides a powerful framework for inferring the statistical properties of unknown parameters that determine the boundary conditions of a physical system. By~incorporating prior knowledge, beliefs, or~assumptions about the values of the parameters, we can obtain more accurate and robust estimates and quantify our uncertainty about the inferred~values.

\subsection{Presentation of the Algorithm for Solving the Inverse~Problem}
\label{inv_algo}

\begin{figure}[H]
\centering
    \includegraphics[scale=0.4]{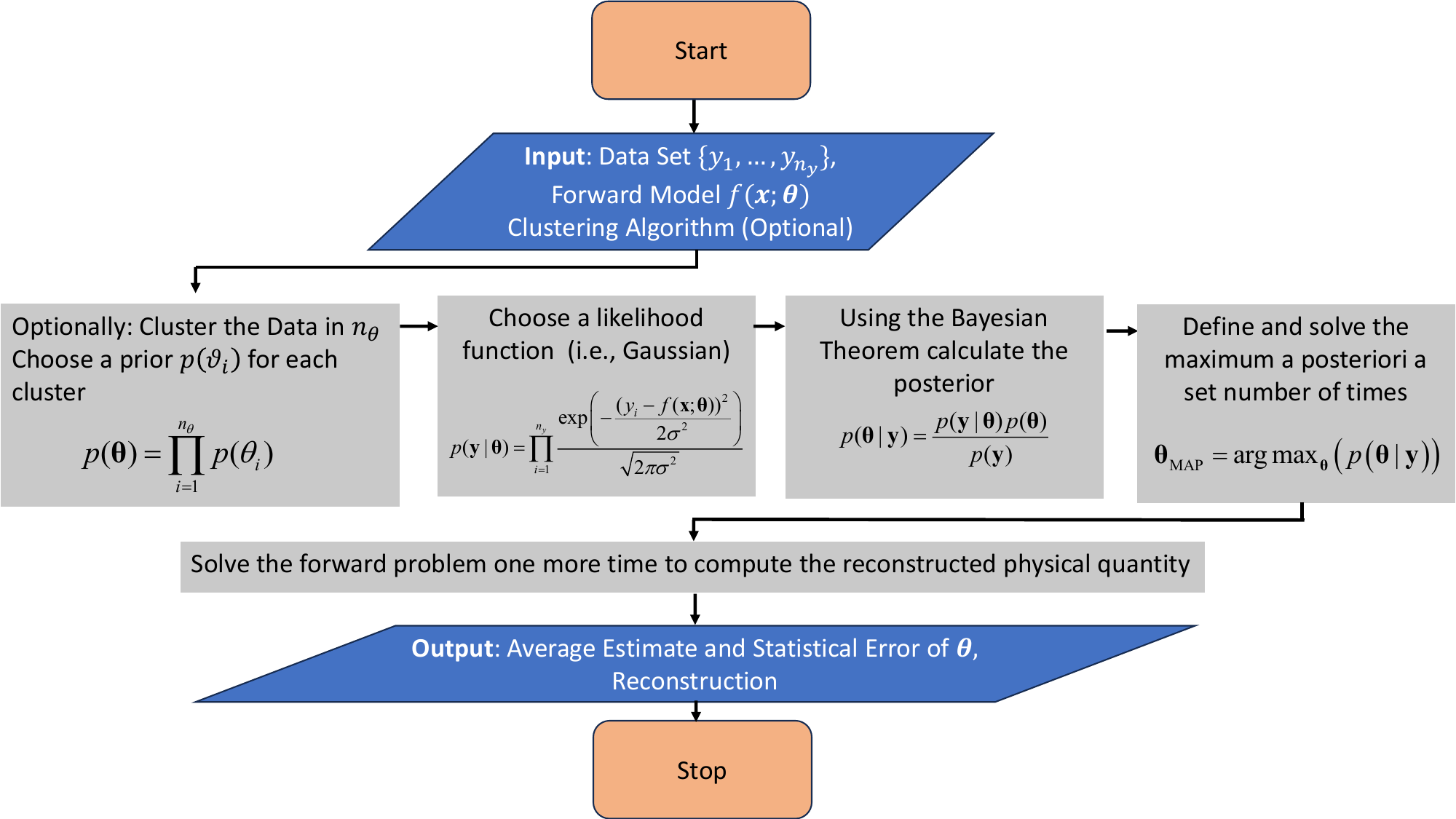}
\caption{Flowchart of our algorithm.}
\label{fig:algorithm}
\end{figure}

The proposed algorithm aims to infer the statistical properties of unknown parameters $\boldsymbol{\theta}$ in a physical system described by differential equations. It is particularly adept at handling inverse problems with uncertain boundary conditions. Here, we present the algorithm (see the flowchart \ref{alg:inverse_problem}) and then discuss its main characteristics and stages (see also the flow diagram in Figure~\ref{fig:algorithm}).

\begin{algorithm}[H]
\caption{Data-driven reconstruction.}
\label{alg:inverse_problem}
\begin{algorithmic}
\Require (Input) Data set $\mathbf{y}$, forward model $f(\mathbf{x};\boldsymbol{\theta})$, noise level $\sigma$, optional clustering algorithm $C(\mathbf{y})$
\Ensure (Output) Estimated value of $\boldsymbol{\theta}$ that maximizes the posterior function, Reconstructed physical quantity
\State \textbf{Step 1:} \textit{Optional Clustering}
    \If{clustering is used}
        \State \hspace{\algorithmicindent}\textbf{(a)} Apply clustering algorithm $C(\mathbf{y})$ to partition $\mathbf{y}$ into clusters
        \State \hspace{\algorithmicindent}\textbf{(b)} For each cluster, select a prior for $\boldsymbol{\theta}$ based on prior knowledge
    \EndIf
\State \textbf{Step 2:} Choose a likelihood function reflecting the probability of observing $\mathbf{y}$ given $\boldsymbol{\theta}$
\State \textbf{Step 3:} Compute the posterior distribution $p(\boldsymbol{\theta} | \mathbf{y})$ via Bayes' theorem
\State \textbf{Step 4:} Define and solve the MAP problem to estimate $\boldsymbol{\theta}$ using numerical optimization, and~repeat a set number of times to refine $\boldsymbol{\theta}$ estimate and assess statistical error
\State \textbf{Step 5:} Solve the forward problem with the estimated $\boldsymbol{\theta}$ to complete reconstruction of the physical quantity in the whole domain
\end{algorithmic}
\end{algorithm}

\vspace{-6pt}

The algorithm comprises several stages, each designed to systematically address different aspects of the inverse problem. Initially, an~optional clustering step groups data points based on their proximity and characteristics, facilitating differentiated analysis for each cluster. This step is particularly useful for sparsely distributed data. The~subsequent steps involve choosing a likelihood function based on the probability of observing the data given the parameters, computing the posterior distribution using Bayes' theorem, and~solving a MAP problem through numerical optimization. This process is iterated to refine the estimates of $\boldsymbol{\theta}$ and assess statistical errors. Finally, the~algorithm solves the forward problem with the estimated parameters, completing the reconstruction of the physical quantity across the entire~domain.

In the following, we provide a few comments on the above reconstruction~algorithm.

\begin{itemize}
\item We assume that the physical system is described by a set of differential equations which involve unknown parameters $\boldsymbol{\theta}$. For~the applications considered here, the~latter appear in the boundary conditions of the system. Therefore, our method can be applied to inverse problems with uncertain boundary conditions, which has received less attention in previous research.
In terms of the physical domain, the~proposed algorithm can be applied to any geometry, regardless of its complexity. The~algorithm can accommodate data that are sparsely distributed across extensive domains.
\item \textcolor{black} {If the data $y$ are sparsely distributed, clustering can be applied to group the data points, enabling a differentiated analysis for each group. Clustering ensures that each region of the domain is assigned a tailored prior distribution, improving the accuracy of the Bayesian inference process. We used the k-means algorithm~\cite{Kaufman1990, Hastie2009} for its simplicity and computational efficiency in our examples, supported by the silhouette method~\cite{Rousseeuw1987}, to determine the optimal number of clusters. Although~k-means is well suited for data with clear boundaries and Euclidean geometry, alternative clustering algorithms (e.g., hierarchical clustering~\cite{Johnson1967}) may be explored in future work. It is important to note that clustering inaccuracies can affect the estimation of boundary conditions and, consequently, the~quality of the reconstruction. To~mitigate this risk, we validated the clustering results using the Silhouette method, ensuring reliable groupings. Additional details are provided in \ref{app_cl}.}
\item Depending on the physical system under consideration, different forward models, likelihood functions, and~prior distributions may be appropriate. Although~normal priors are assumed in this paper, alternative distributions, such as exponential or gamma, may be more suitable for other problems.
\textcolor{black} {\item The optimization algorithm chosen to maximize the posterior function must be in line with the complexity and accuracy requirements of the problem at hand. For~this work, we employed stochastic optimization methods~\cite{Maroufpoor2020}, specifically dual annealing~\cite{Xiang1997} and differential evolution~\cite{Lampinen}, due to their robustness in handling non-convex optimization problems. These methods are particularly effective in navigating high-dimensional parameter spaces and avoiding local minima, a~critical consideration given the complexity of the posterior distributions in our Bayesian framework. In~addition, these approaches reduce the risk of bias associated with initial guessing, ensuring a more thorough exploration of the solution space. For~more information on optimization algorithms, see  \ref{app3}.}
\item We consider Gaussian noise for our likelihood function and a constant noise \mbox{term $\sigma$.} However, the~noise level can also be treated as a random parameter within the Bayesian framework, allowing us to estimate the noise in conjunction with the \mbox{model parameters.}
\item Our algorithm is constructed to operate within a complete Bayesian framework. It can accommodate clustering if necessary and, despite the increased computational demand, is designed to solve the forward model directly.
\item Surrogate models~\cite{Kennedy2001} can be used to reduce computational costs but can affect precision depending on the quality of the approximation.
\end{itemize}

\textcolor{black}{
Overall, the~proposed algorithm is versatile and adaptable to various domains, accommodating different conditions and data distributions. It provides a comprehensive framework for exploring the statistical properties of unknown quantities within a full Bayesian context. A~key innovation of our approach lies in its ability to handle sparse and irregularly distributed data in large domains while remaining robust to noise in observations. By~integrating clustering, Bayesian inference, and~numerical simulations, the~algorithm is uniquely equipped to partition observational data into tailored priors, ensuring accurate and region-specific reconstructions even in challenging scenarios. 
Compared to existing methods, such as Markov chain Monte Carlo (MCMC)-based approaches, our algorithm achieves comparable accuracy while significantly reducing computational overhead through the use of maximum a posteriori (MAP) estimation. Additionally, its reliance on stochastic optimization methods, such as dual annealing and differential evolution, ensures robustness against local minima and facilitates exploration of non-convex parameter spaces. Unlike physics-informed machine learning methods that often require extensive datasets and high computational resources, our algorithm performs effectively with sparse data, making it both data-efficient and scalable. 
This versatility and efficiency highlight the potential of our approach for applications in diverse fields, including astrophysics, geophysics, and~engineering. In~Section~\ref{inv_gmf}, we will demonstrate the application of the algorithm to reconstruct a magnetic field from synthetic sparse data that simulate actual physical measurements. Further comparisons with state-of-the-art techniques, including MCMC algorithms, will be explored in a forthcoming study. A~discussion of the space and time complexity of the algorithm is presented in  \ref{complexity}.}

\textcolor{black} {In our examples in the rest of this work, we choose a normal distribution as the prior due to its simplicity, symmetry, and~mathematical convenience in Bayesian inference, particularly when limited information about the parameter distributions is available. Its continuous nature and central tendency make it well-suited for modeling expected variations in boundary conditions, aligning with the physical assumptions of our \mbox{inverse problem.}}

\textcolor{black}{The application presented in this work, the~reconstruction of the magnetic field (MF) in a cone domain using sparse data, can be considered a first step towards the reconstruction of the galactic magnetic field (GMF). This is a critical problem in astrophysics with significant implications for understanding the structure and evolution of the Milky Way~\cite{Tsouros2024}. Among~other physical effects, the~GMF influences the dynamics of cosmic rays~\cite{Magkos2019} and plays a role in star formation processes~\cite{Tassis2004}. A~detailed mapping of the galactic MF would offer transformative insights into these astrophysical phenomena. However, observations of the GMF are often sparse, noisy, and~unevenly distributed across the sky, making the inverse problem of magnetic field reconstruction highly ill-posed~\cite{Pelgrims2022}. Current state-of-the-art studies mainly rely on best-fit methods~\cite{Jansson2012,Jansson2012a}, which can be limited in their ability to capture the full complexity of the magnetic field structure. In~contrast, the~methodology presented in this work aims to address this challenge by employing a Bayesian inference framework, where the magnetic field is reconstructed from limited data by combining physical constraints and statistical modeling. Although~the current study uses synthetic data sets for controlled testing and validation, the~ultimate goal is to extend this approach to real astronomical observations once a sufficient volume of direct measurements becomes available~\cite{Panopoulou2019}. This future step will provide a more realistic and rigorous test of the method’s ability to recover the structure of the GMF under observational constraints.}

\section{The Forward~Problem}
\label{gmf}
\unskip

\subsection{The Finite Element Method in the Unconstrained~Problem}
\subsubsection{Mathematical~Details}
\label{uncon}
Here, we present the forward problem that governs MF reconstruction. We compute the MF (denoted as $\mathbf{B}$) numerically in a proper 3D geometry given specific boundary conditions with the finite element method (FEM) \cite{Huebner2001, Logan2012}, by~dividing our system into smaller and simpler parts, the~finite elements. To~implement our partial differential \mbox{equation (PDE)} problem in numerical code, we use the popular {\it FEniCSx} Python package~\cite{Alnaes2014, Scroggs2022a, Scroggs2022}. The~MF is considered time-independent, and there are no sources, so the Maxwell equation system reduces to the solution of a Laplace equation, also imposing the constraint $\nabla \cdot \mathbf{B}=0$. We approach the forward problem in two variants: an unconstrained scenario where the Poisson equation is solved directly and~a constrained scenario where the solution is subject to the divergence-free condition of the Magnetic~Field. 

In the first scenario, we consider the Poisson PDE with a source term ${\bm{\rho }} (\bf{x})$ (even though, in our application, we have no sources we keep the discussion here quite general. Note also that we assume that the MF is time-independent) and accompanied with boundary conditions (BC) of the Dirichlet type:
\begin{equation}
\label{PDE}
\begin{split}
- { \nabla ^2}\mathbf{B}(\mathbf{x}) & = \bm{\rho} (\mathbf{x}),~~\mathbf{x} \in \Omega, \\
\mathbf{B}(\mathbf{x}) & = \mathbf{f}(\mathbf{x}),~~\mathbf{x} \in \partial \Omega.
\end{split}
\end{equation}

We use the Cartesian coordinate system with ${\bf{x}} = (x,y,z)$ as the coordinate vector. Our computational domain $\Omega$ or mesh or geometry of our problem will be a proper 3D object like a cone, and $\partial \Omega$ is its'~boundary.

This problem is equivalent to the functional minimization of the following action $J$:
\begin{equation}
\label{un}
J(\mathbf{B})=\frac{1}{2}\int_{\Omega}{d\mathbf{x}}\left| \nabla \mathbf{B} \right|^2-\int_{\Omega}{d\mathbf{x}}{\bm{\rho }}\cdot \mathbf{B},
\end{equation}
where $\nabla \mathbf{B}$ is the gradient of a vector. 

For a general vector, say $\mathbf{u}$,  $\nabla \mathbf{u}$ is its' gradient, i.e.,~a matrix/tensor ($\boldsymbol{e}_{\boldsymbol{i}}\boldsymbol{e}_{\boldsymbol{j}}\frac{\partial u_j}{\partial x_i}$ in Cartesian notation) or simply the Jacobean. So, 

\begin{equation}
\left| \nabla \mathbf{u} \right|^2=\mathrm{tr}\left( \left( \nabla \mathbf{u} \right) ^2 \right) =\sum_{j=1}^3{\frac{\partial u_j}{\partial x_j}\frac{\partial u_j}{\partial x_j}}=\left( \frac{\partial u_x}{\partial x} \right) ^2+\left( \frac{\partial u_y}{\partial y} \right) ^2+\left( \frac{\partial u_z}{\partial z} \right) ^2.
\end{equation}

The FEM is applied to convert the PDE for our physical system into a variational problem~\cite{Huebner2001, Logan2012}. To~achieve this, we multiply the PDE with a test function $\upsilon^k(\mathbf{x})$, which vanishes at the boundary and integrates over the domain $\Omega$:

\begin{equation}
\label{variational0}
-\int\limits_{\Omega}{d\mathbf{x}}\left( \nabla ^2B^k(\mathbf{x}) \right) \upsilon ^k(\mathbf{x})=\int\limits_{\Omega}{d\mathbf{x}}\rho ^k(\mathbf{x})\upsilon ^k(\mathbf{x}),~~\mathbf{x}\in \Omega,~~k=x,y,z.
\end{equation}

In FEM terminology, $\boldsymbol{\upsilon }(\mathbf{x})\equiv \left( \upsilon ^x(\mathbf{x}),\upsilon ^y(\mathbf{x}),\upsilon ^z(\mathbf{x}) \right) $ is called a test function and lies in a vector function space $\hat{V}$, while the unknown function $\bf{B}(\bf{x})$ is called a trial function and lies in a possible different vector function space $V$. Integration by parts on the first-hand side of this equation gives us the following equation (the surface term that is also produced vanishes because ${\bm \upsilon} ({\bm x})$ vanishes in the boundary):
\begin{equation}
\label{variational}
\int\limits_{\Omega}{d\mathbf{x}}\nabla B^k(\mathbf{x})\cdot \nabla \upsilon ^k(\mathbf{x})=\int\limits_{\Omega}{d\mathbf{x}}\rho ^k(\mathbf{x})\upsilon ^k(\mathbf{x}),~~\mathbf{x}\in \Omega,~~k=x,y,z.
\end{equation}

This is the variational or weak form of our original problem (see Equation \eqref{PDE}). The~trial and test spaces in this problem are the following:
\begin{equation}
\label{trial1}
V=\left\{ H^1\left( \Omega \right) :\boldsymbol{\upsilon }(\mathbf{x})=\mathbf{f}(\mathbf{x})~\mathrm{for}~\mathbf{x}\in \partial \Omega \right\}, 
\end{equation}
and
\begin{equation}
\label{test1}
\hat{V}=\left\{ H^1\left( \Omega \right) :\boldsymbol{\upsilon }(\mathbf{x})=0~\mathrm{for}~\mathbf{x}\in \partial \Omega \right\}.
\end{equation}

In the above equations, $H^1\left( \Omega \right) $ is the Sobolev space containing functions $\boldsymbol{\upsilon }$, so both $\boldsymbol{\upsilon }$ and $\nabla \boldsymbol{\upsilon }$ are square integrable. To~solve the problem of Equation \eqref{variational}, we have to transform it from a continuous variational problem to a discrete one by introducing finite-dimensional test and trial spaces $V_h\subset V$ and $\hat{V}_h\subset \hat{V}$, respectively, and~the choice of the said spaces follows directly from the finite elements we want to use in our solution. Assume that we have a basis $\left\{ \phi _j \right\} _{j=1}^{N}$ for $V_h$ and a basis $\left\{ {\hat{\phi}} _j \right\} _{j=1}^{N}$ for $\hat{V}_h$ where $N$ denotes the dimension of the spaces mentioned above, which is equal to the number of nodes in the FEM scheme. Although~the analytical form of the basis depends on the actual finite element, we use the fact that the value of ${\phi} _j$ is one in the $j$ node and zero in the others. If~we denote the coordinates of each node as $\mathbf p_i$; then,
\begin{equation}
\label{basis}
\phi _j\left( \mathbf{p}_i \right) =\delta _{ji}=\left\{ \begin{array}{c}
	1\mathrm{ },\mathrm{ }j=i\\
	0\mathrm{ },\mathrm{ }j\ne i\\
\end{array}, \right. 
\end{equation}
where $\delta _{ji}$ is the Kronecker~delta. 

We now make an ansatz for discretizing ${\bf{B}}$, denoted as ${\bf{B}}_h \in V_h$:
\begin{equation}
\label{ansatz}
\begin{split}
\mathbf{B}_h=\left( \sum_{j=1}^N{U_{j}^{x}\phi _j},\sum_{j=1}^N{U_{j}^{y}\phi _j}\sum_{j=1}^N{U_{j}^{z}\phi _j} \right),~~U_{j}^{x,y,z}=B_{h}^{x,y,z}\left( \mathbf{p}_j \right) ,~~\mathbf{p}_j\in \mathrm{nodes},
\end{split}
\end{equation}
where, for each component, we split the sum in the inner and boundary nodes as follows:
\begin{equation}
\label{ansatz2}
\begin{split}
\sum_{j=1}^N{U_{j}^{k}\phi _j}=\sum_{j=1}^{N_{\mathrm{in}}}{U_{j}^{k}\phi _j}+\sum_{j=1}^{N_{\mathrm{bn}}}{U_{j}^{k}\phi _j},k=x,y,z,
\end{split}
\end{equation}
where $\left( U_{j}^{x},U_{j}^{y},U_{j}^{z} \right)\in \mathbb{R} ^{3N}$ is the vector of degrees of freedom to be computed and we split the sum in the inner and boundary nodes, with~$N_{\mathrm{in}}$ being the number of the former and $N_{\mathrm{bn}}$ the number of the latter ($N_{\mathrm{in}}+N_{\mathrm{bn}}=N$).  However, from~the Dirichlet boundary conditions, we can immediately find the unknowns $U_{j}^{k}$ that correspond to the \mbox{boundary nodes:}
\begin{equation}
B_{h}^{k}\left( \mathbf{p}_j \right) =f^k\left( \mathbf{p}_j \right) =U_{j}^{k},\mathbf{p}_j\in \mathrm{boundary}~~ \mathrm{nodes} ~~k=x,y,z..
\end{equation}

Finally, if~we choose
\begin{equation}
\boldsymbol{\upsilon }=\left( \phi _i,\phi _i,\phi _i \right) ,~~~i=1,N_{\mathrm{in}}
\end{equation}
from Equation \eqref{variational}, we can assemble the following system of algebraic equations:
\begin{equation}
\label{variational2}
\begin{split}
\sum_{j=1}^{N_{\mathrm{in}}}{U_{j}^{k}\int\limits_{\Omega}{d\mathbf{x}}\nabla \phi _j}\cdot \nabla \phi _i=\int\limits_\Omega  {d{\bf{x}}} {\rho ^k}({\bf{x}}){\phi _i}-\sum_{j=1}^{N_{\mathrm{bn}}}{f_{j}^{k}\int\limits_{\Omega}{d\mathbf{x}}\nabla \phi _j}\cdot \nabla \phi _i,~~i=1,N_{\mathrm{in}},~~k=x,y,z.
\end{split}
\end{equation}

The resulting algebraic system includes the stiffness matrix derived from the inner products of basis function gradients:
\begin{equation}
\label{stiff}
A_{ij}=\int\limits_{\Omega}{d\mathbf{x}}\nabla \phi _i\cdot \nabla \phi _j.
\end{equation}

The solution of the above system provides the coefficients $U_j^k$, thus obtaining the \mbox{solution $\mathbf{B}_h$} throughout the domain.
There are many ways to solve Equation~(\ref{variational2}) numerically. In~our case, we opted to use a direct solver (LU-factorization) \cite{trefethen1997}.

\subsubsection{The Geometry and Finite Element Method~Basis}
\label{fembasis}

\begin{figure}[H]
\begin{subfigure}[b]{0.45\textwidth}
\includegraphics[scale=0.45]{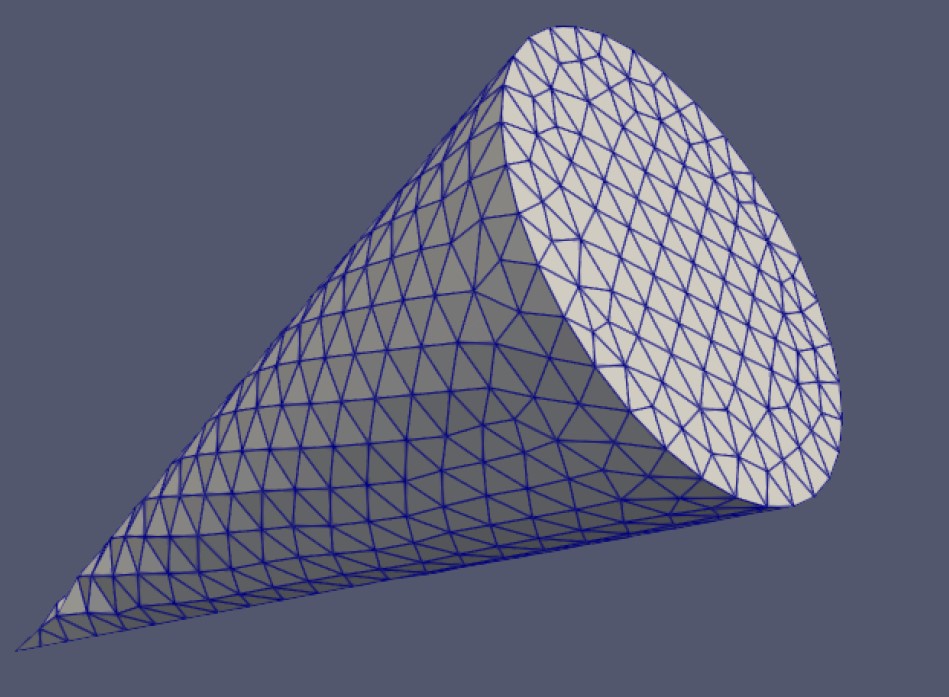}
\caption{\centering}
\label{fig:cone1}
\end{subfigure}
\hfill
\begin{subfigure}[b]{0.45\textwidth}
\centering 
\includegraphics[scale=0.55]{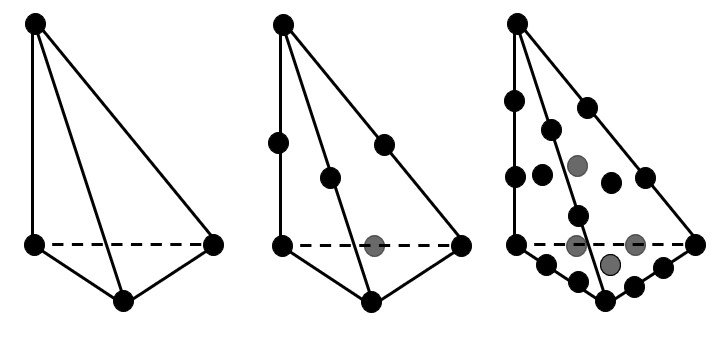}
\caption{\centering}
\label{fig:pyramids}
\end{subfigure}
    
\caption{The physical--computational domain of our problem (left figure) and the finite elements used (right figure). (\textbf{a}) An example of our cone~geometry. (\textbf{b}) The finite elements that we use. From~left to right: $\mathbb{P}_1$, $\mathbb{P}_2$, and $\mathbb{P}_3$ triangle~pyramids.}
\label{fig:3and4}
\end{figure}

For the geometric representation of our cone domain, we use a mesh divided into tetrahedral elements (see Figure~\ref{fig:3and4}a): we divide the cone into triangular pyramids~\cite{Logg2012a}. The~solution will be calculated as points in the pyramids (nodes). We will consider \mbox{three cases} for the elements: $\mathbb{P}_1$, where the nodes are in the vertices; $\mathbb{P}_2$, where the nodes are on the vertices and in the center of each side of the triangles; $\mathbb{P}_3$, where they are on the vertices, with one on each third or each side and one in the center of each triangle (see Figure~\ref{fig:3and4}b). These elements are also called Lagrange elements~\cite{Logg2012a, Arnold2014, Cockburn2017}. In~our case, vector-valued (or generally tensor-valued) Lagrange elements are constructed using a Lagrange element for \mbox{each component.}

\vspace{-6pt}

For the basis that we use in our calculation, see \cite{Logg2012a, defelement}. For~example, the~basis functions for a $\mathbb{P}_1$ tetrahedron with nodes at the points $(0,0,0)$, $(1,0,0)$, $(0,1,0)$, $(0,0,1)$ are as follows:
\begin{equation}
\label{basisP1}
\left\{ \begin{array}{c}
	\phi _0=1-x-y-z\\
	\phi _1=x\\
	\phi _2=y\\
	\phi _3=z\\
\end{array} \right\}. 
\end{equation}

This basis respects the property of Equation \eqref{basis}. Before~substituting it into \mbox{Equation \eqref{ansatz}} or \eqref{variational2}, we have to apply a coordinate transformation to account for the position and orientation of the tetrahedron in the physical~mesh. 

In cases of systems involving multiple PDEs, such as a vector and a scalar function, a~mixed function space is employed, with~the popular Taylor--Hood element \(\mathbb{P}_k-\mathbb{P}_{k-1}\) being a common choice for such scenarios. This element allows for a distinct approximation of vector and scalar fields, ensuring accuracy and compliance with the constraints of governing~equations.

\subsubsection{Example}
\label{examples}

For the boundary conditions $\mathbf{f}(\mathbf{x})\equiv \mathbf{B}_{\mathrm{BC}}(\mathbf{x})$ we generally consider two cases, including a~purely analytical one with boundary conditions with respect to equation $\nabla  \cdot {\bf{B}}=0$, for~example,
\begin{equation}
\label{bound_un_1}
\mathbf{B}_{\mathrm{BC}}(\mathbf{x})=\left( 10x+y-z,~~~x-15y+z,~~~x-y+5z \right) ~\mathrm{for}~\mathbf{x}=\left( x,y,z \right) \in \partial \Omega.
\end{equation}

\begin{figure}[H]
\centering
    \begin{subfigure}[b]{0.4\textwidth}
        \centering
        \includegraphics[width=\textwidth]{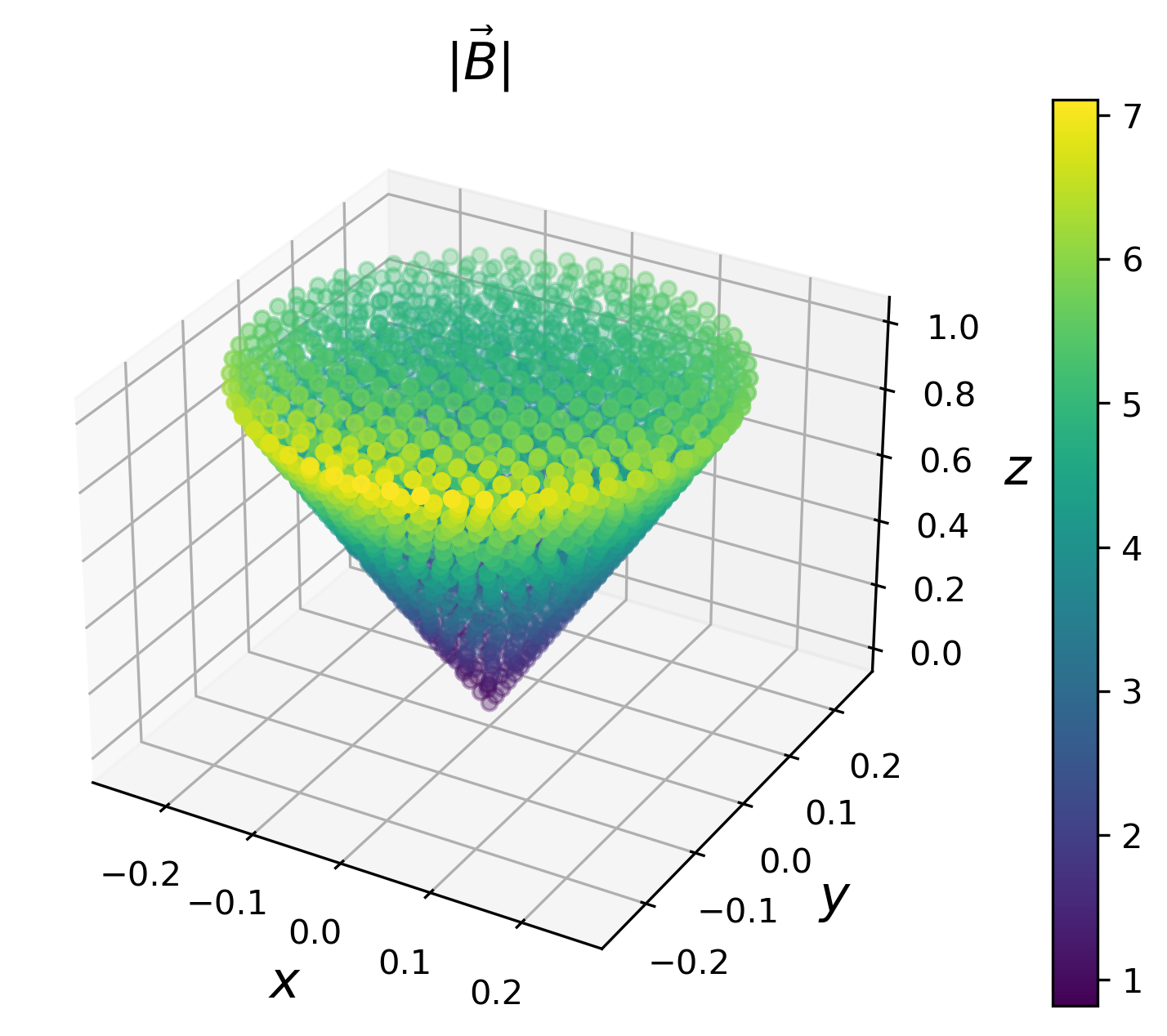}
        \caption{\centering}
        \label{fig:image1}
    \end{subfigure}
    \begin{subfigure}[b]{0.4\textwidth}
        \includegraphics[width=\textwidth]{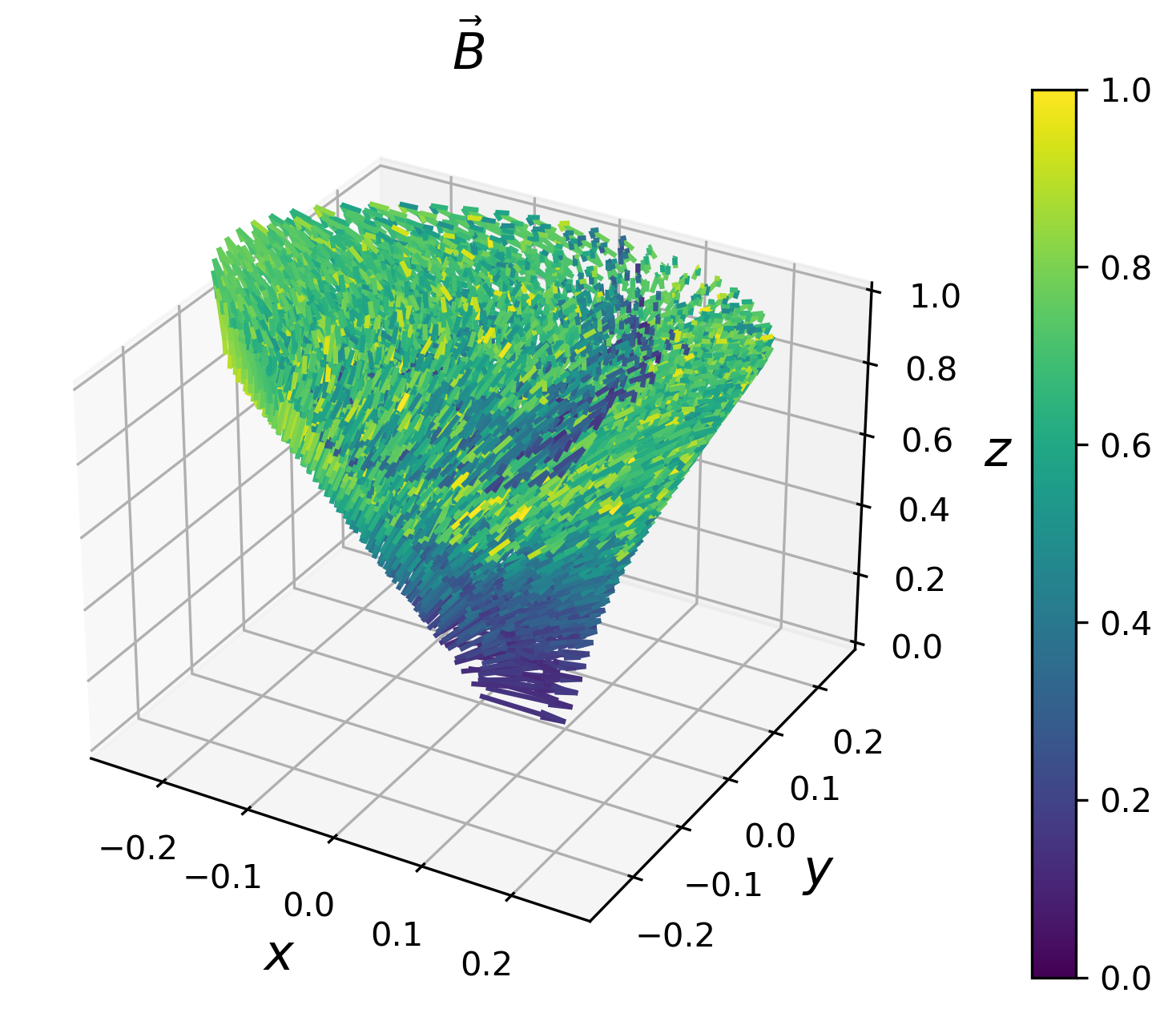}
        \caption{\centering}
        \label{fig:image2}
    \end{subfigure}

    \vspace{0.5cm} 

    \begin{subfigure}[b]{0.4\textwidth}
        \includegraphics[width=\textwidth]{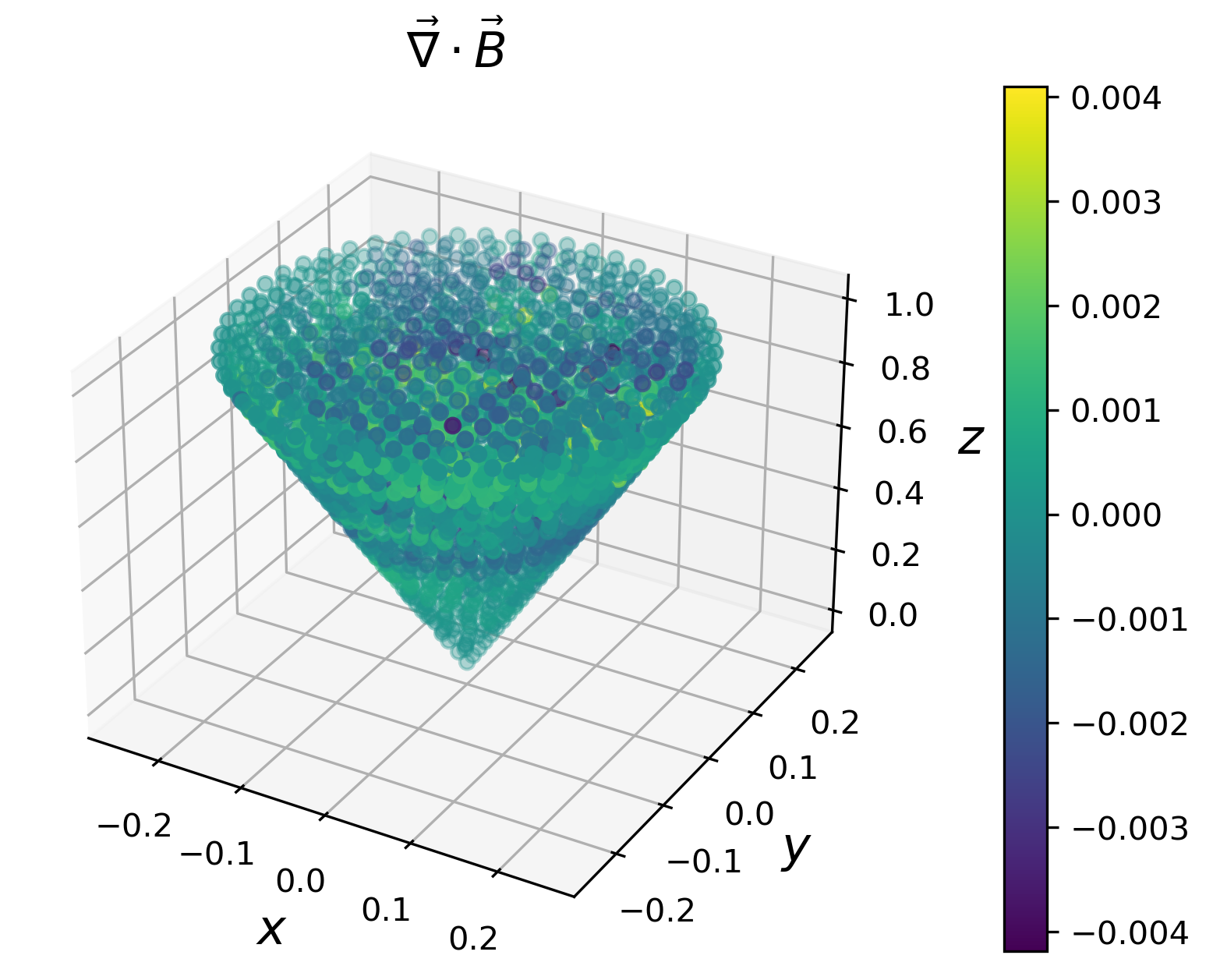}
        \caption{\centering}
        \label{fig:image3}
    \end{subfigure}
    \begin{subfigure}[b]{0.4\textwidth}
        \includegraphics[width=\textwidth]{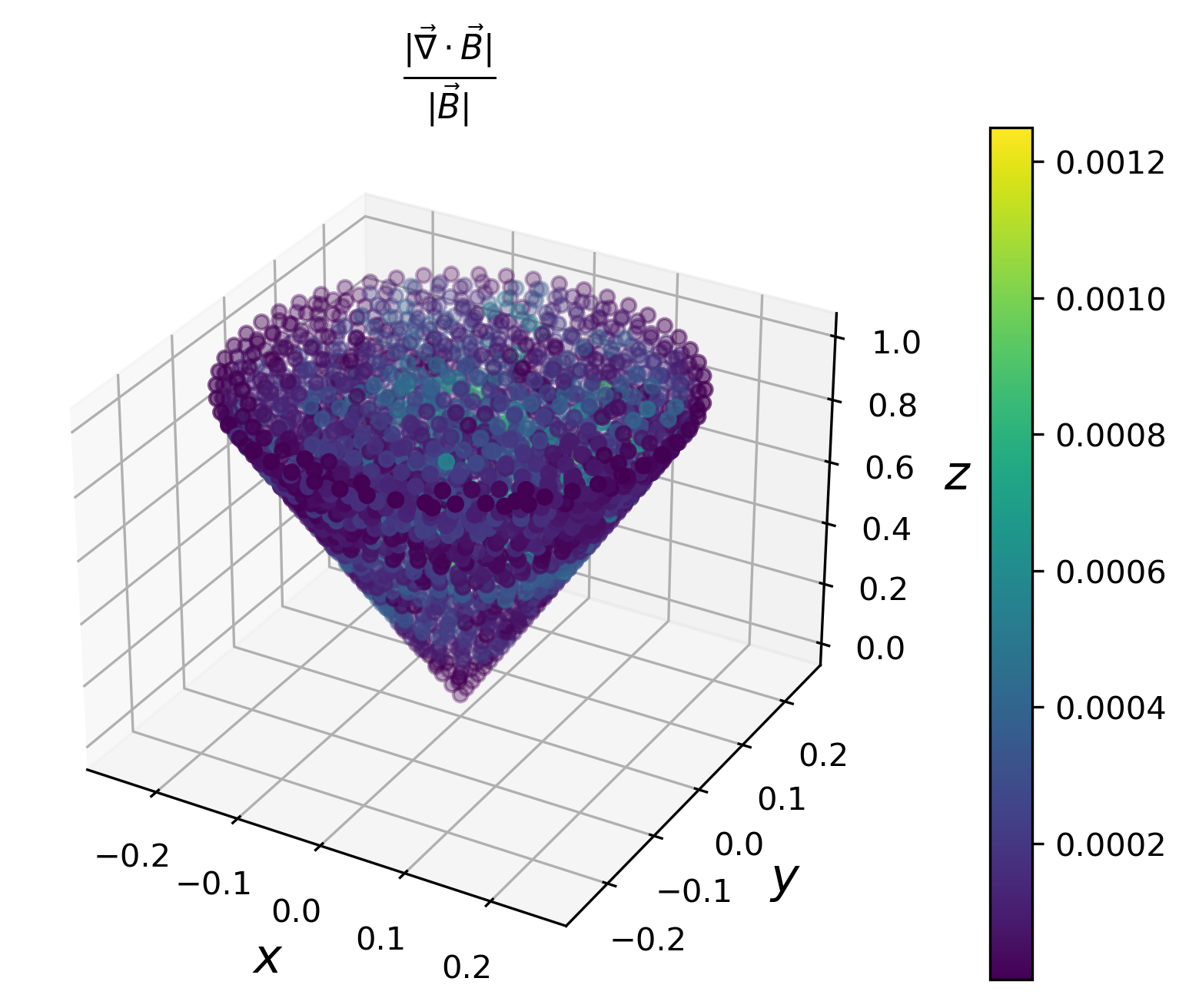}
        \caption{\centering}
        \label{fig:image4}
    \end{subfigure}


    \caption{Calculation of MF for the unconstrained case (${\nabla ^2}{\bf{B}}({\bf{x}}) = 0$) with analytical BC calculated from Equation \eqref{bound_un_1} in a 3D cone domain. We plot the magnitude of the MF (\textbf{a}), its vector (\textbf{b}), its divergence (\textbf{c}), and~the ratio of the magnitude of the divergence to the magnitude of the MF (\textbf{d}). We can see that $\nabla  \cdot {\bf{B}} \sim {10^{ - 3}}$.}
    \label{fig:bound_un_1}
\end{figure}

We solve the problem in a cone geometry with its vertex at origin $(0,0,0)$, a height of $1$, and base radius of $0.25$ using the finite element $\mathbb{P}_1$. The~results are shown in Figure~\ref{fig:bound_un_1}a,b. This finite element was also used to check the accuracy of our solver in cases that can be solved analytically. The~results shown in Figures \ref{fig:bound_un_1}a and \ref{fig:bound_un_1}b illustrate the magnitude and vector direction of the MF, respectively. Furthermore, Figures \ref{fig:bound_un_1}c and \ref{fig:bound_un_1}d provide information on the divergence of the MF and the relationship between the absolute value of the divergence and the magnitude of the MF. The~color gradients in these figures correspond to different magnitudes of the MF, with~the divergence nearing $\nabla \cdot {\bf{B}} \sim {10^{ - 3}}$, indicating a low divergence, as~expected from the boundary condition that enforces $\nabla \cdot {\bf{B}}=0$. 
However, as~we will see in the next example, this value can be reduced further if we consider a constrained calculation where the constraint $\nabla \cdot {\bf{B}}=0$ is taken into account in the FEM~scheme.

\vspace{-9pt}

Another case of interest is if one or more boundary conditions are generated from one of more normal distributions ${\cal N}(m,s)$, where $m$ is the mean value and $s$ is the standard deviation of each distribution. However, this calculation will be presented in Section~\ref{inv_gmf}, when we discuss the generation of synthetic sparse~data.

\subsection{Constrained~Minimization}
\unskip

\subsubsection{Formulation of the Forward Constrained~Problem}
\label{fotfcp}

For the solution to represent an actual magnetic field. we have to ensure that our solution obeys Maxwell's equation $\nabla  \cdot {\bf{B}}=0$. This means that we have to solve the Poisson equation while ensuring that the aforementioned condition holds.
The above problem is equivalent to the functional minimization of the following action $J$:
\begin{equation}
\label{lagra}
J(\mathbf{B},\lambda )=\frac{1}{2}\int_{\Omega}{d\mathbf{x}}\left| \nabla \mathbf{B} \right|^2-\int_{\Omega}{d\mathbf{x}}\boldsymbol{\rho }\cdot \mathbf{B}+\int_{\Omega}{d\mathbf{x}\lambda \nabla \cdot \mathbf{B}},
\end{equation}
where $\lambda$ is a Lagrange multiplier. Minimizing with respect to $\bf{B}$ gives
\begin{equation}
\label{eq1}
-\nabla ^2\mathbf{B}(\mathbf{x})=\nabla \lambda +\boldsymbol{\rho }(\mathbf{x}),~~\mathbf{x}\in \Omega,
\end{equation}
while minimization with respect to $\lambda$ gives back the following constraint:
\begin{equation}
\label{eq2}
\nabla  \cdot {\mathbf{B}}({\bf{x}}) = 0. 
\end{equation}

The rigorous proof is shown in  \ref{motcl}.

This problem is similar to the simple steady-state Navier--Stokes problem~\cite{Gresho2000}:
\begin{equation}
\begin{array}{l}
-\nabla ^2\mathbf{u}-\nabla P=\boldsymbol{\rho } \\
~~\nabla \cdot \mathbf{u}=0
\end{array},
\end{equation}
accompanied, of~course, by~proper boundary conditions. In~this equation, $\mathbf{u}$ is the velocity and pressure $P$ plays the role of the Lagrange~multiplier.

\subsubsection{Finite Element Method Applied in the Constrained Minimization~Problem}

From the results of the previous section, we can summarize the PDE problem \mbox{as follows:}
\begin{equation}
\label{constrained}
\begin{split}
- { \nabla ^2}\mathbf{B}(\mathbf{x}) & = \nabla \lambda(\mathbf{x}) + \bm{\rho}(\mathbf{x}),~~\mathbf{x} \in \Omega, \\
\nabla \cdot \mathbf{B} & = 0,~~\mathbf{x} \in \Omega, \\
\mathbf{B}(\mathbf{x}) & = \mathbf{f}(\mathbf{x}),~~\mathbf{x} \in \partial \Omega.
\end{split}
\end{equation}

The boundary condition for the Lagrange multiplier $\lambda$ can be found from the first PDE of this system up to a constant. To~solve the problem with FEM, we need to find the variational form of the above system. To~define the test and trial functions, we use a mixed functional space. Consider the vector function spaces \eqref{trial1} and \eqref{test1} and the function space:
\begin{equation}
\label{trial2}
Q=\left\{q\in L^2(\Omega)\left|\; \int_{\Omega}{d\mathbf{x}}q=0\right.\right\},
\end{equation}
where $L^2$ is the space of square-integrable functions. Our solution is that $(\mathbf{B}, \lambda) \in W $. The~space $W$ is a mixed (product) function space $W = V \times Q$, such that $\mathbf{B} \in V$ and $\lambda \in Q$. In~a similar manner for the trial functions, we have $(\boldsymbol{\upsilon },q)\in \hat{W}= \hat{V}\times Q$. We multiply the first PDE of the system shown in Equation \eqref{constrained} with the test function $\boldsymbol{\upsilon }$ and the second one with the test function $q$. Similarly to Section~\ref{uncon}, we have the following:
\begin{equation}
\label{convar1}
\begin{split}
\int\limits_{\Omega}{d\mathbf{x}}\nabla B^k(\mathbf{x})\cdot \nabla \upsilon ^k(\mathbf{x})+\int\limits_{\Omega}{d\mathbf{x}}\lambda(\mathbf{x}) \partial _k\upsilon ^k(\mathbf{x}) & =\int\limits_{\Omega}{d\mathbf{x}}\rho ^k(\mathbf{x})\upsilon ^k(\mathbf{x}),~~k=x,y,z\\
\int_{\Omega}{d\mathbf{x}}\nabla \cdot \mathbf{B}(\mathbf{x})q(\mathbf{x}) & =0,~~\mathbf{x}\in \Omega.
\end{split}
\end{equation}

The ansatz for the magnetic field ${\bf{B}}_h \in V_h$ is the same as that of the unconstrained case and is given in Equation \eqref{ansatz}. In~a similar manner, for~the Lagrange multiplier, we consider a basis $\left\{ \varphi _j \right\} _{j=1}^{M}$ for $Q_k\subset Q$, and the relevant expansion is as follows:
\begin{equation}
\label{ansatzl}
\lambda =\sum_{j=1}^M{L_{j}\varphi _j}. 
\end{equation}

If we substitute this equation in the system \eqref{convar1} and set $\upsilon ^k=\phi _i,~ i=1,N$ and $q=\varphi _i,~ i=1,M$, we find the following:
\begin{equation}
\label{convar3}
\begin{split}
\sum_{j=1}^N{A_{ij}U_{j}^{k}}+\sum_{j=1}^M{C_{ij}L_j} & =\int\limits_\Omega  {d{\bf{x}}} {\rho ^k}({\bf{x}}){\phi _i},~~~i=1,N,~~~k=x,y,z \\
\sum_{k=x,y,z}{\sum_{j=1}^N{C_{ji}^{k}U_{j}^{k}}}&=0,~~~i=1,M,
\end{split}
\end{equation}
where the $N \times N$ stiffness matrix $\mathbf{A}=\left\{ A_{ij} \right\}, ~i,j=1,N $ is given by Equation \eqref{stiff}, and we have also introduced the following $N \times M$ matrix:
\begin{equation}
\label{bb}
\mathbf{C}=\left\{ \begin{array}{c}
	C_{ij}^{x}\\
	C_{ij}^{y}\\
	C_{ij}^{z}\\
\end{array} \right\} ,~~i=1,\ldots,N,~~j=1,\ldots,M,
\end{equation}
where
\begin{equation}
\label{bbb}
C_{ij}^{k}=\int\limits_{\Omega}{d\mathbf{x}}\partial _k\phi _i\varphi _j.
\end{equation}

So, if~we assemble the system in matrix form, we have the following:
\begin{equation}
\label{convar4}
\left[ \begin{matrix}
	\mathbf{A}&		\mathbf{0}&		\mathbf{0}&		\mathbf{C}^{\mathbf{x}}\\
	\mathbf{0}&		\mathbf{A}&		\mathbf{0}&		\mathbf{C}^{\mathbf{y}}\\
	\mathbf{0}&		\mathbf{0}&		\mathbf{A}&		\mathbf{C}^{\mathbf{z}}\\
	\left( \mathbf{C}^{\mathbf{x}} \right) ^{\bot}&		\left( \mathbf{C}^{\mathbf{y}} \right) ^{\bot}&		\left( \mathbf{C}^{\mathbf{z}} \right) ^{\bot}&		\mathbf{0}\\
\end{matrix} \right] \left[ \begin{array}{c}
	\mathbf{U}^{\mathbf{x}}\\
	\mathbf{U}^{\mathbf{y}}\\
	\mathbf{U}^{\mathbf{z}}\\
	\mathbf{L}\\
\end{array} \right] =\left[ \begin{array}{c}
	\left\{ \rho _{i}^{x}\phi _i \right\}\\
	\left\{ \rho _{i}^{y}\phi _i \right\}\\
	\left\{ \rho _{i}^{z}\phi _i \right\}\\
	\mathbf{0}\\
\end{array} \right], 
\end{equation}
where $\mathbf{C}^k=\left\{ C_{ij}^{k} \right\} ,\mathbf{U}^k=\left\{ U_{j}^{k} \right\} ,~k=x,y,z,~i=1,\ldots,N,~j=1,\ldots,M$.

For finite elements, we opted to use the Taylor--Hood ones that we mentioned in Section~\ref{fembasis}, namely the $\mathbb{P}_3-\mathbb{P}_{2}$ one. So, in~the expansions of each component of $\bf{B}(\bf{x})$, we use the $\mathbb{P}_3$ tetrahedron, and~in the Lagrange multiplier expansion, we use the $\mathbb{P}_2$ tetrahedron. This finite element is a popular choice in the solution of the Navier--Stokes problem~\cite{John2017}; however, the~construction of finite elements that are divergence-free is the subject of active research~\cite{Sime2021}.

The system \eqref{convar4} is a linear system of the saddle-point type~\cite{Benzi2005}. To~solve it, we use a preconditioner that generally transforms a linear system into another system with more favorable properties for linear solvers. For~finite element problems, a usual choice would be a block diagonal matrix defined as $\mathbf{P}=\mathrm{diag}\left\{ \mathbf{A},\mathbf{A},\mathbf{A},\mathbf{M} \right\} $, where $\mathbf{M}$ is a mass matrix \mbox{with the following elements:}
\begin{equation}
\label{mass2}
M_{ij}=\int\limits_{\Omega}{d\mathbf{x}}\phi _i\varphi _j.
\end{equation}

After applying the above transformation, the new system is solved with the minimal residual method~\cite{Paige1975}.

\subsubsection{Example}

\begin{figure}[H]
    \centering
    \begin{subfigure}[b]{0.4\textwidth}
        \centering
        \includegraphics[width=\textwidth]{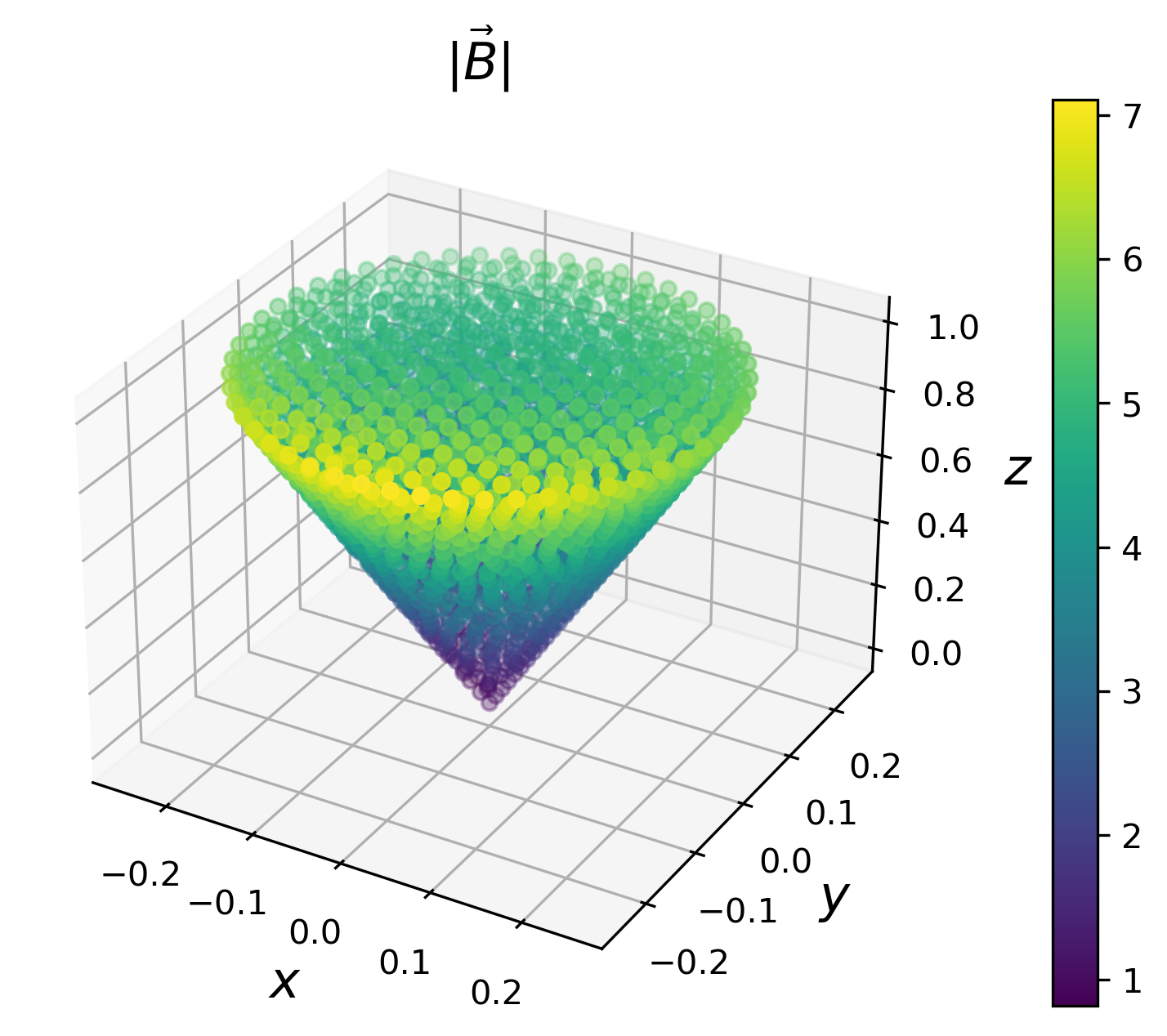}
        \caption{\centering}
        \label{fig:image1a}
    \end{subfigure}
    \begin{subfigure}[b]{0.4\textwidth}
        \centering
        \includegraphics[width=\textwidth]{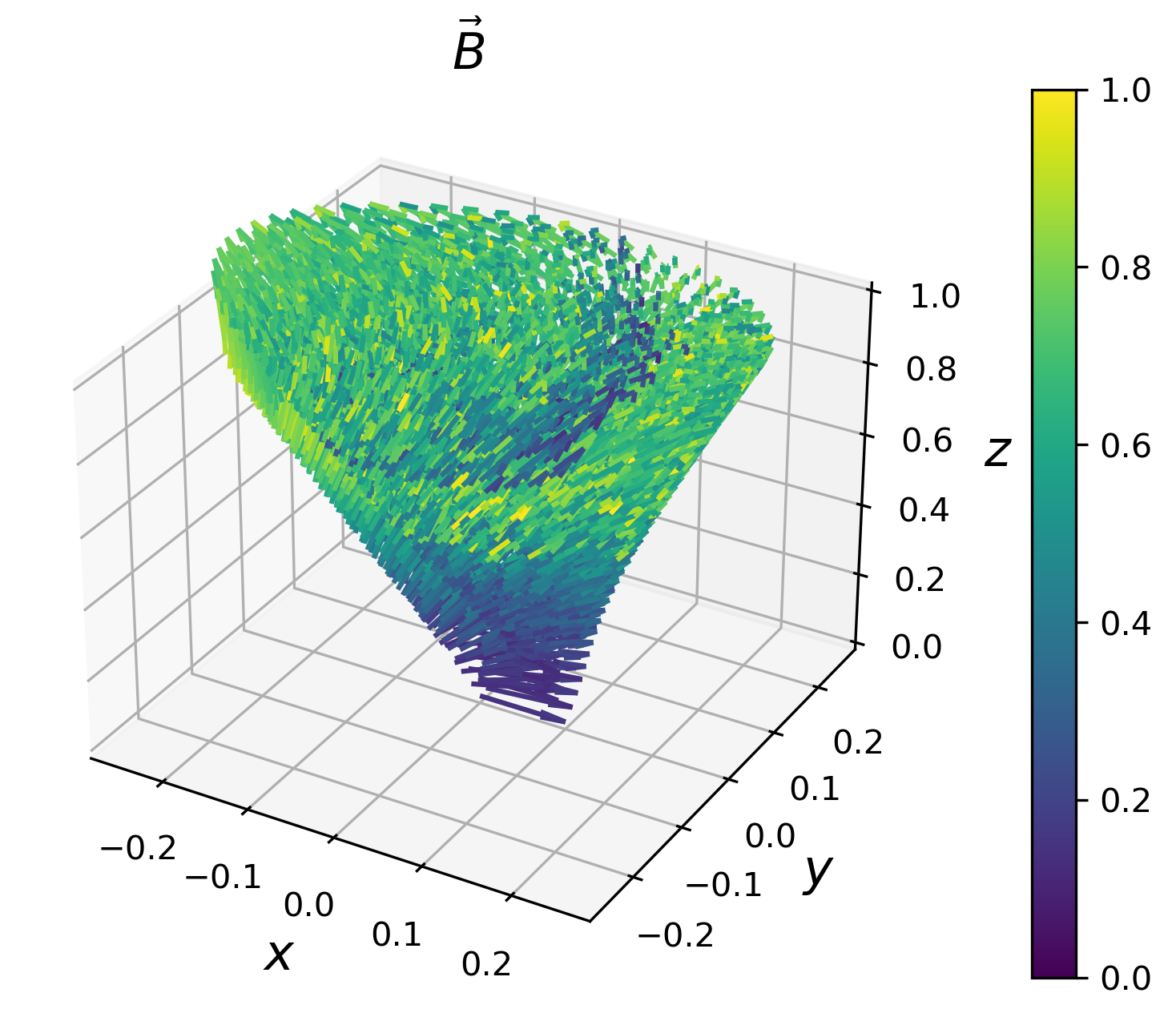}
        \caption{\centering}
        \label{fig:image2a}
    \end{subfigure}
    
    \vspace{0.4cm}
    
    \begin{subfigure}[b]{0.4\textwidth}
        \includegraphics[width=\textwidth]{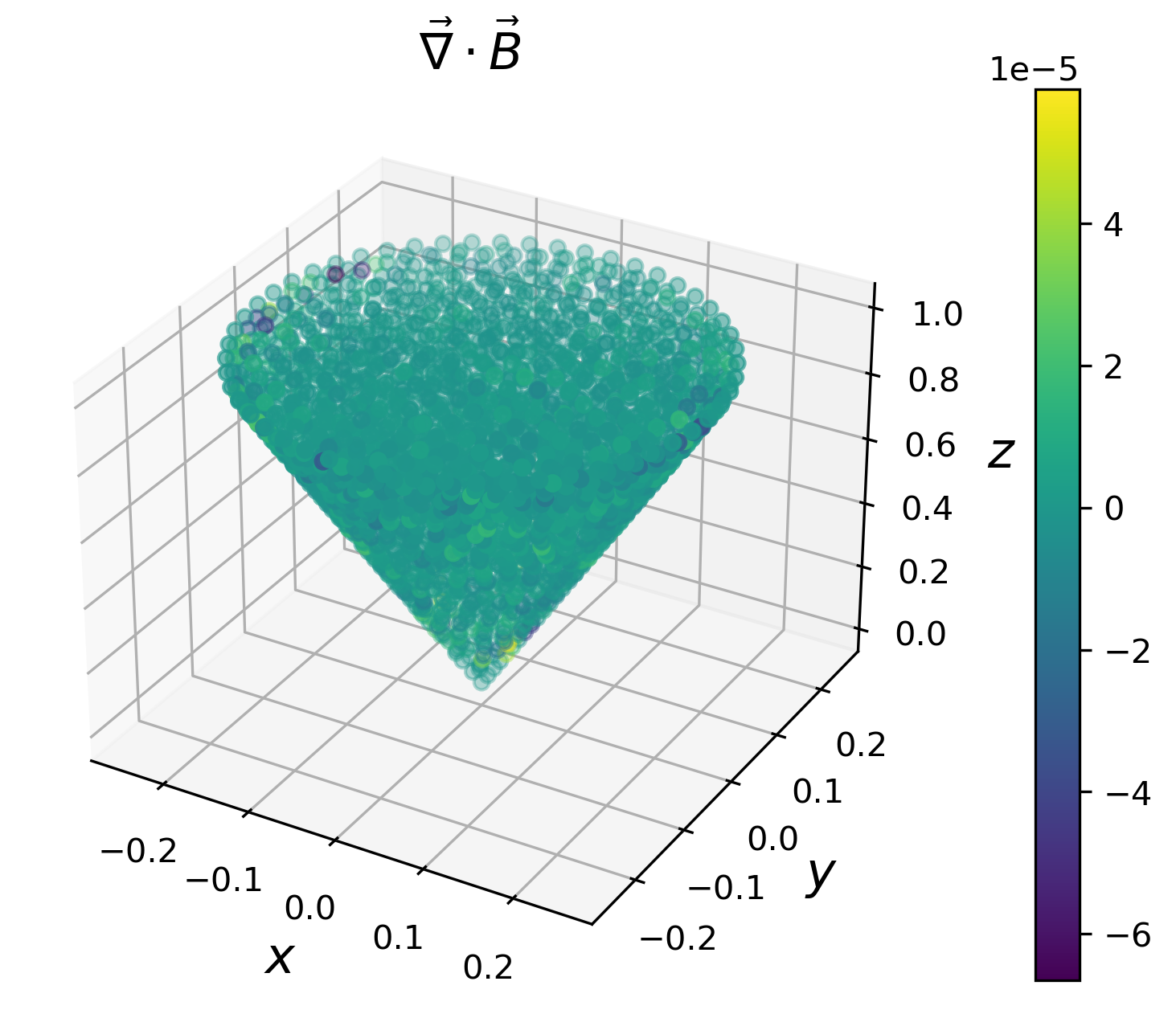}
        \caption{\centering}
        \label{fig:image3a}
    \end{subfigure}
    \begin{subfigure}[b]{0.4\textwidth}
        \centering
        \includegraphics[width=\textwidth]{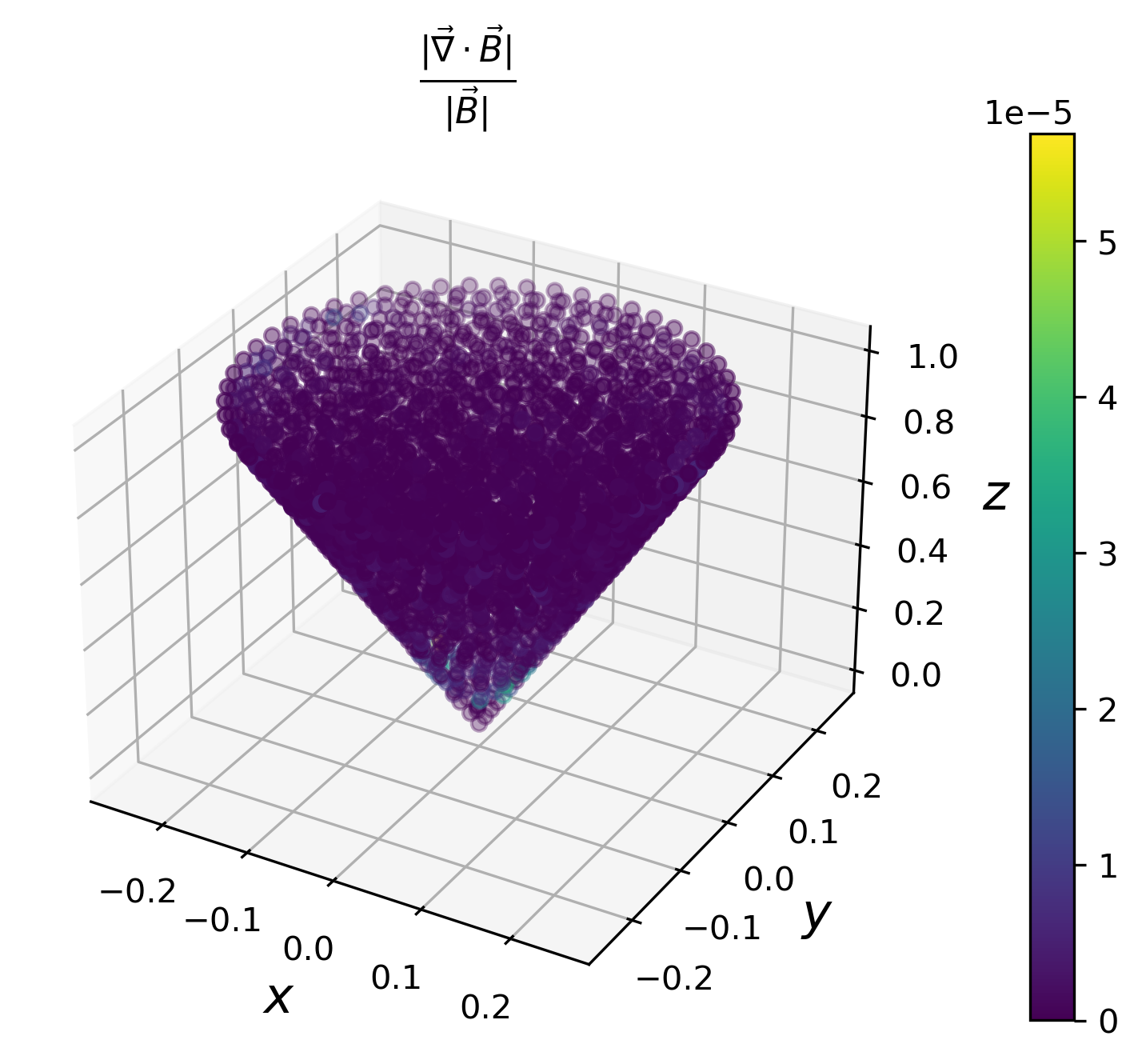}
        \caption{\centering}
        \label{fig:image4a}
    \end{subfigure}
    
    \caption{Calculation of MF for the constrained case (where we impose the constraint $\nabla  \cdot {\bf{B}} = 0$) with analytical BC calculated from Equation~(\ref{bound_un_1}) in a 3D cone domain. We plot the magnitude of the MF (\textbf{a}), its vector (\textbf{b}), its divergence (\textbf{c}), and the ratio of the magnitude of the divergence to the magnitude of the MF (\textbf{d}). We can see that $\nabla  \cdot {\bf{B}} \sim {10^{ - 5}}$, which is an improvement over what we can see in Figure~\ref{fig:bound_un_1}.}
    \label{fig:bound_con_1}
\end{figure}

In this example, we revisit the calculations from Section~\ref{examples}, this time incorporating the divergence-free constraint $\nabla \cdot {\bf{B}} =0$ as a fundamental aspect of our model from the outset, using the Taylor--Hood element $\mathbb{P}_3-\mathbb{P}_{2}$. The~results, illustrated in Figure~\ref{fig:bound_con_1}, demonstrate the impact of this constraint on the~calculation.

\vspace{-6pt}

Comparing the constrained results with those of the unconstrained case \mbox{(Figure \ref{fig:bound_un_1}),} we observe a marked improvement in adherence to the divergence-free condition. Figures \ref{fig:bound_con_1}c and \ref{fig:bound_con_1}d show a significant reduction in divergence values across the domain, confirming the effectiveness of the constraint in preserving the physical precision of our model. This example underscores the importance of incorporating such constraints into computational models to ensure that the results are not only mathematically robust but also physically meaningful.

\section{Reconstructing the Magnetic Field in a Specific~Domain}
\label{inv_gmf}
\unskip

\subsection{The Reconstruction~Process}
\label{app_alg_gmf}

In this section, we present an application of our proposed algorithm. Using synthetic data and the solution of the forward problem, we reconstruct the MF in an arbitrarily given domain. To~simulate the sparse direct observational data of the MF, we start by creating synthetic data from a simulation using the forward problem in the cone domain $\Omega$ for the following boundary conditions. 
As a first example, we consider the following \mbox{boundary conditions:}
\begin{equation}
\label{bound_un_2}
\mathbf{B}_{\mathrm{BC}}(\mathbf{x})=\left( \begin{matrix}
	{\cal N}(m,s), ~~~2y-5z, ~~~10y-2z\\
\end{matrix} \right)~~\mathrm{for}~\mathbf{x}=\left(x,y,z\right)\in \partial \Omega.
\end{equation}

In this equation, the normal distribution \({\cal N}(m, s)\) is equivalent to $f(\mathbf{x} ; \theta) + \epsilon$, as~seen in Equation \eqref{first}, provided \(\theta = m\), with~\(\epsilon \sim {\cal N}(0, s)\). In~the following, since in the actual physical problem, we may not know the true boundary conditions, but rather only have access to uncertain measurements, we choose to use BCs as above, that is, consisting of one or more normal distributions for the x-component. So, as~shown in Equation \eqref{bound_un_2}, our first application will be with a normal distribution with mean $m=10$ and standard deviation $s=0.5$. We will also consider the case where the cone is divided into planes, and we have a different normal distribution generating the BC in each one. Although~we focus on normal distributions for the x-component in this case, our approach is capable of handling BCs derived from any probability density function (PDF), catering to the diverse and complex nature of actual measurement~methods.

Having solved the forward problem and obtained a set of data to~simulate irregularly distributed sparse data, we randomly remove a percentage of the points from the aforementioned set of data. The~remaining points represent the sparse direct observational data of the x component of the MF that we will use in the application of our algorithm. It is important to note that for simplicity, we only used normal distributions for the x component of the boundary conditions in this simulation. In~a more realistic scenario, the~above scenario may be applied to all components of the MF or, alternatively, separately to the strength and direction of the~field.

We can now proceed with the application of the inverse problem algorithm, as~was introduced in Section~\ref{inv_algo}. The~unknown parameters $\theta_i$ are the mean values of the \mbox{x-component} of the MF at the boundary of the cone while the input is the simulated synthetic sparse data for the x-component of the MF, say $y_i, ~~i=1,\ldots,n_y$, the~forward problem, $f(\mathbf{x} ; \boldsymbol{\theta})$, which was analyzed in Section~\ref{gmf} (in this application we consider the unconstrained mode), the~noise term that characterizes the measurement error (we choose $\sigma=1$) and a clustering~algorithm.

In order to maintain a clear focus on the main contributions of the article while providing full transparency of our method, the~detailed steps of the example application of our algorithm are presented in \ref{application}. This includes a step-by-step account of the clustering approach, the~definition and minimization of the cost function (which is the weighted sum of the differences between the data set values minus the result of the forward problem), the~optimization process, and~the calculation of residuals, culminating in the reconstruction of the magnetic field within the entire domain of interest. These supplementary steps exemplify the robustness and applicability of our algorithm in various scenarios and ensure that interested readers can replicate or build on our work. Here, we only focus on the general key points: we have one prior for each possible cluster, and we consider each to be a normal one; therefore, the~total prior is their product. We also consider Gaussian likelihood probability. The~likelihood and posterior functions are defined from Equations \eqref{likelihood} and \eqref{Bayes}, respectively, while the maximum a posteriori problem is defined from Equation \eqref{mle}.

Our proposed inverse problem algorithm should then recover the mean values of these distributions. So, for~example, if~the synthetic data were created using Equation \eqref{bound_un_2}, we should find, for~$\theta$, a~value close to 10. A~solution of the forward model gives the MF in the entire cone domain, thus completing the reconstruction~calculation.

We move on to different test cases demonstrating the problem of reconstructing the magnetic field within a given domain while also elaborating on the computational efficiency of our~algorithm.

\subsection{Test Case 1: Single Prior~Data}

\begin{figure}[H]
\centering
    \begin{subfigure}[b]{0.44\textwidth}
        \includegraphics[width=\textwidth]{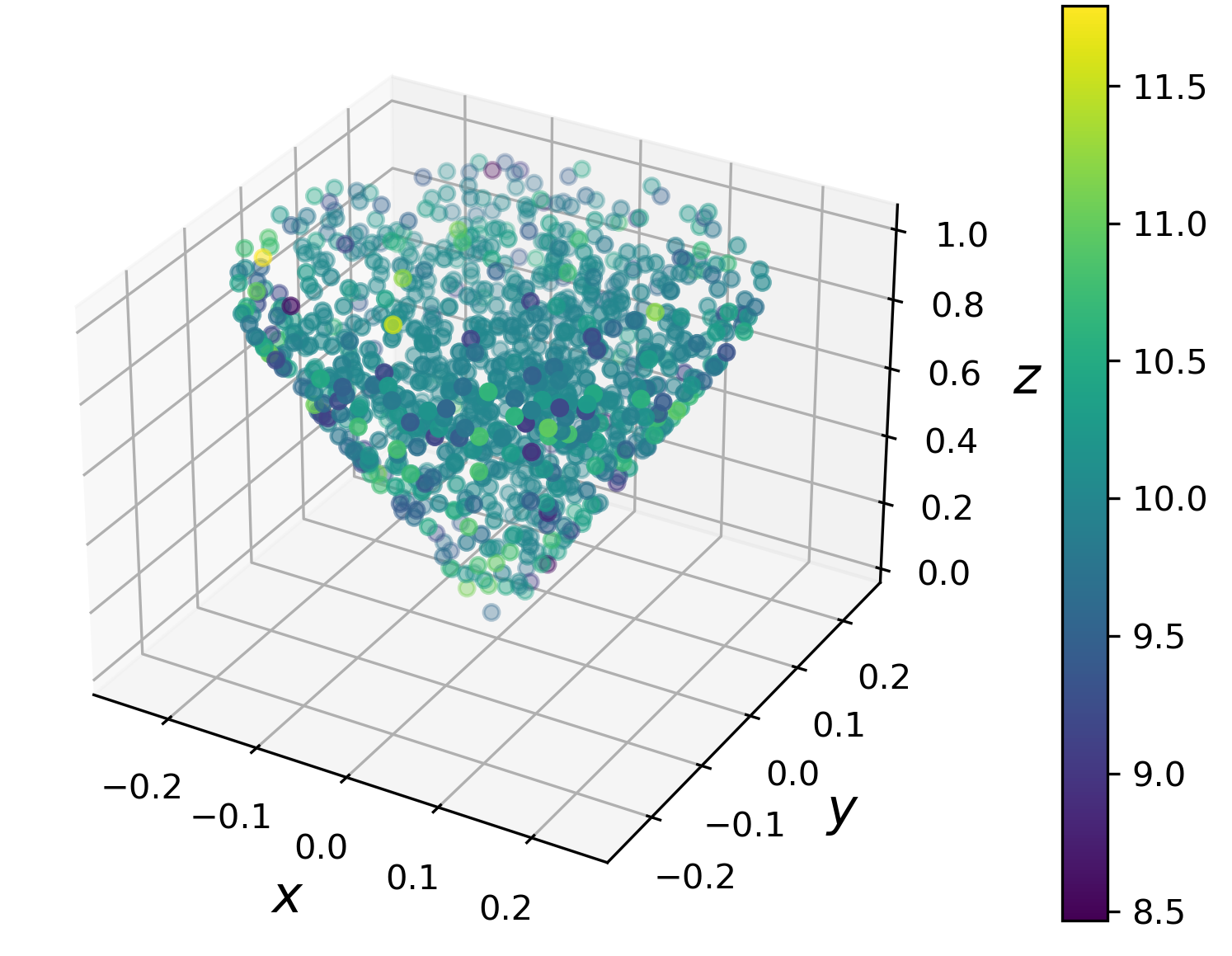}
        \caption{\centering}
        \label{fig:image1b}
    \end{subfigure}
    \hspace{10pt}
    \begin{subfigure}[b]{0.44\textwidth}
        \includegraphics[width=\textwidth]{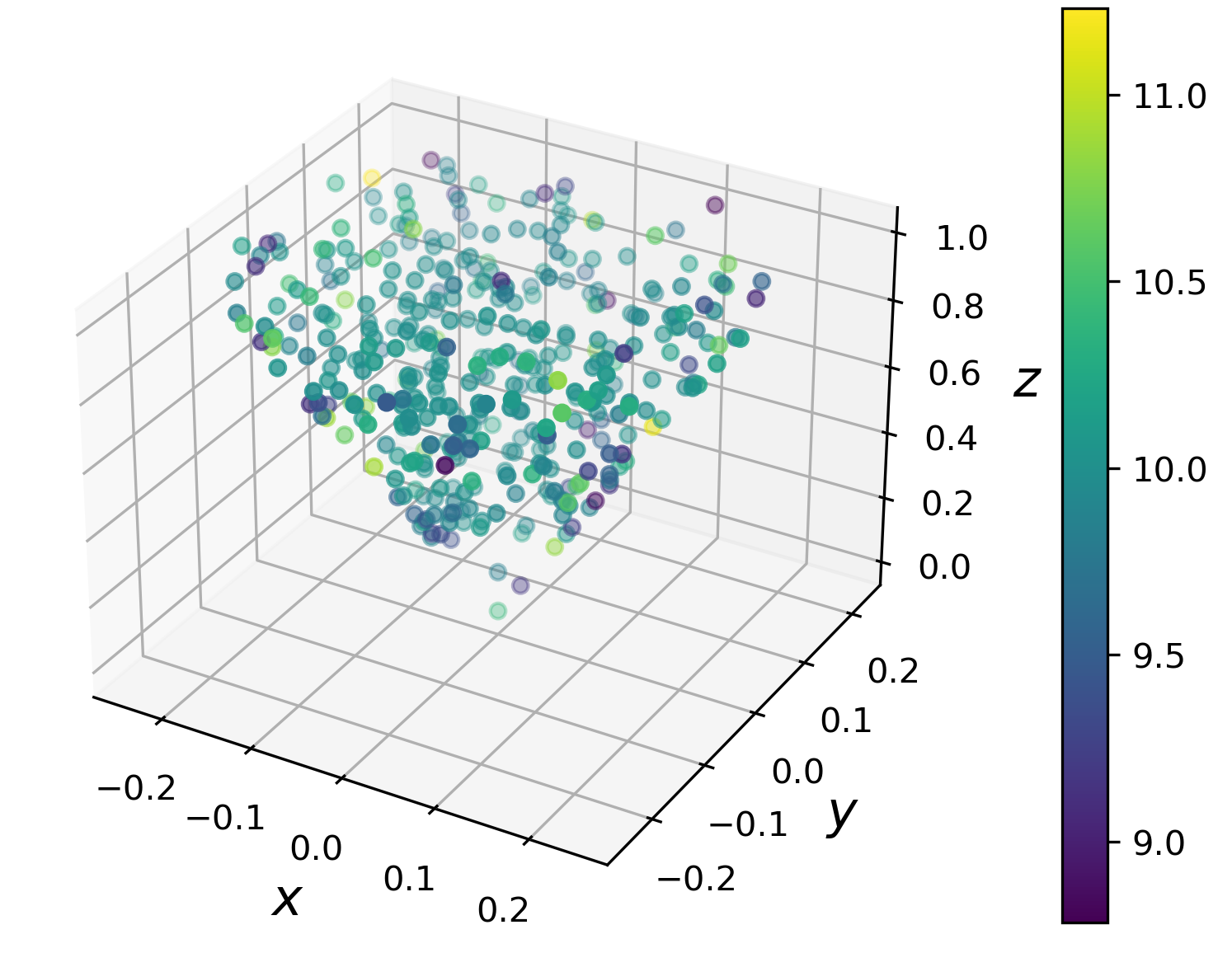}
        \caption{\centering}
        \label{fig:image2b}
    \end{subfigure}
    \vskip\baselineskip
    \begin{subfigure}[b]{0.44\textwidth}
        \includegraphics[width=\textwidth]{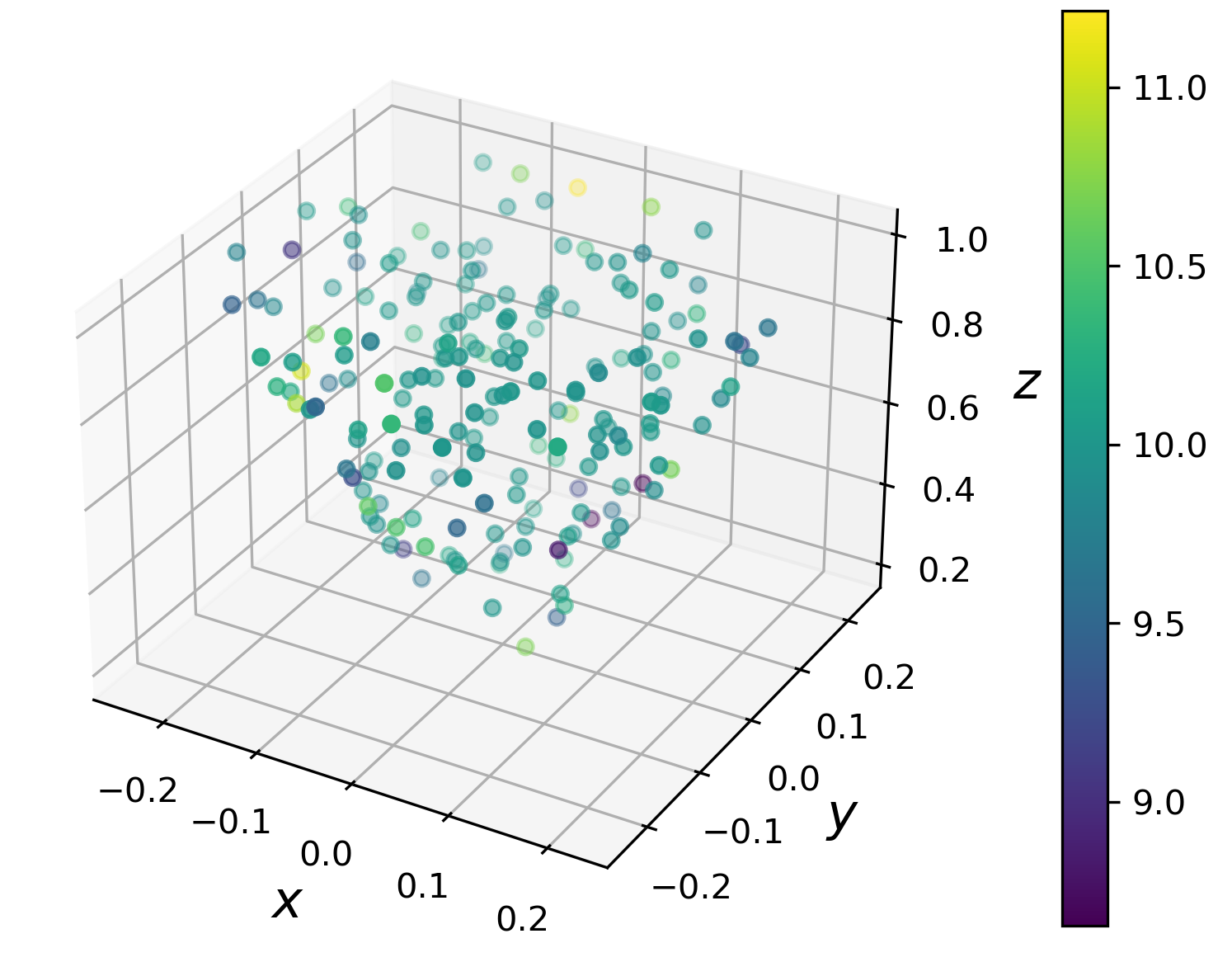}
        \caption{\centering}
        \label{fig:image3b}
    \end{subfigure}
    \hspace{10pt}
    \begin{subfigure}[b]{0.44\textwidth}
        \includegraphics[width=\textwidth]{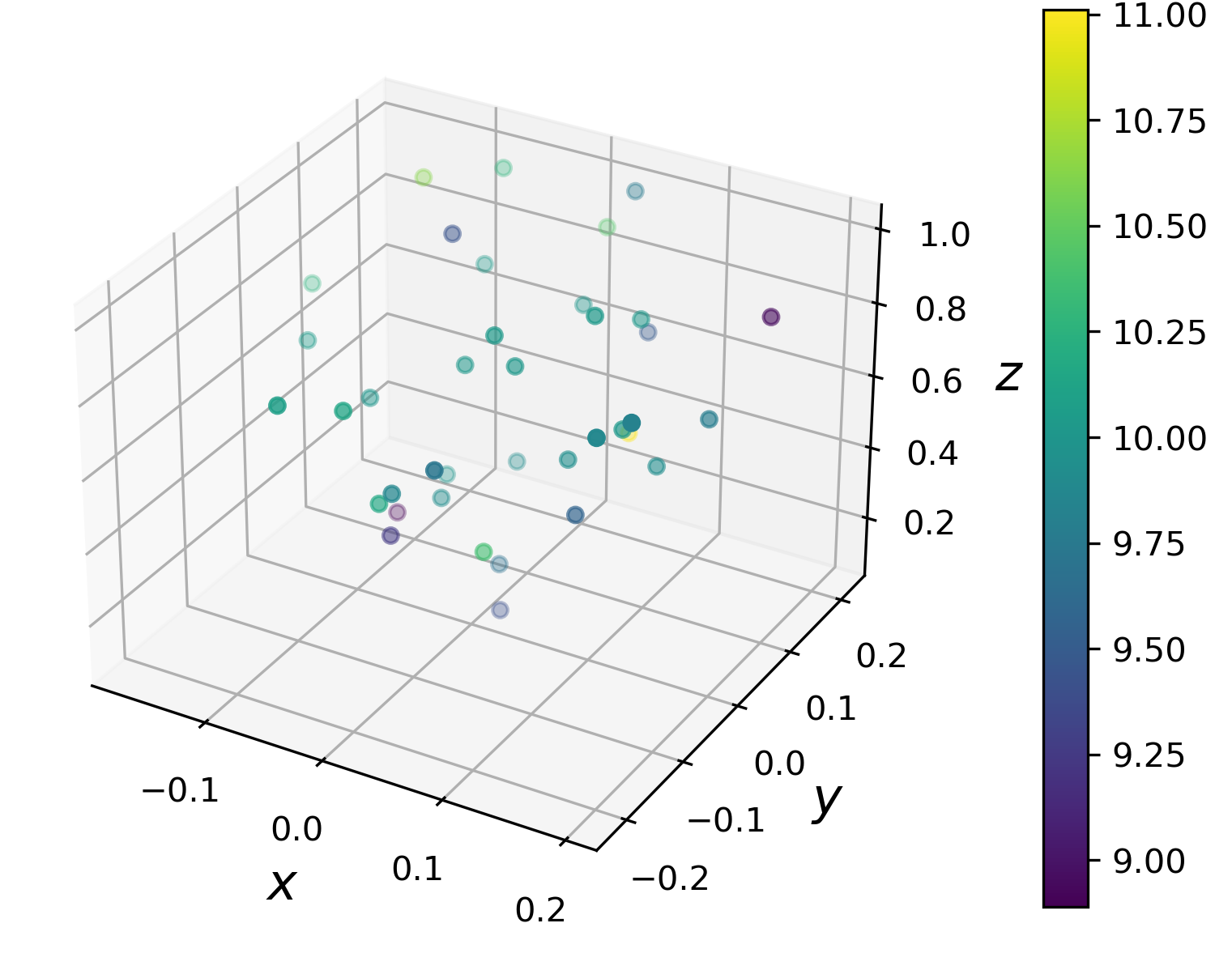}
        \caption{\centering}
        \label{fig:image4b}
    \end{subfigure}
        \vspace{6pt}
    \caption{Synthetic data of the x-component of the MF to be used in the data-driven reconstruction calculation, as calculated from the unconstrained case forward model with the BCs from Equation~(\ref{bound_un_2}). In~addition, in~each sub-figure, a~different percentage of the data has been randomly removed to simulate a case of ``reconstruction from sparse~data''. (\textbf{a}) $75\%$ of the data~removed. (\textbf{b}) $90\%$ of the data~removed. (\textbf{c}) $95\%$ of the data~removed. (\textbf{d}) $99\%$ of the data~removed.}
    \label{fig:sparse1}
\end{figure}

In our first example, we focus on boundary conditions (BCs) for the x-component of the MF generated by a single distribution, as~detailed in Equation~\eqref{bound_un_2} and~employing an unconstrained forward model (see Section~\ref{uncon}). The~domain of interest is a cone with its apex at the origin $(0,0,0)$, a~height of $1$, and a base radius of $0.25$. After~trying various finite elements, we chose the $\mathbb{P}_1$ ones since they give accurate results in this problem. When solving the forward problem within this geometry, we simulate sparse data by randomly omitting a percentage of the approximately 3000 data points within the domain, as~shown in Figure~\ref{fig:sparse1}.

The synthetic data visualized in Figure~\ref{fig:sparse1} illustrates various degrees of data sparsity, ranging from $75\%$ to $99\%$ of random data removal. These subfigures effectively demonstrate the progression of data reduction and its implications for the reconstruction process. In~each case, our aim was to reconstruct the x-component of the MF from these increasingly sparse datasets, and~we applied our algorithm four times, each for each data set shown in the aforementioned~subfigures.

\begin{figure}[H]
    \begin{subfigure}[b]{0.45\textwidth}
\centering 
\includegraphics[width=\textwidth]{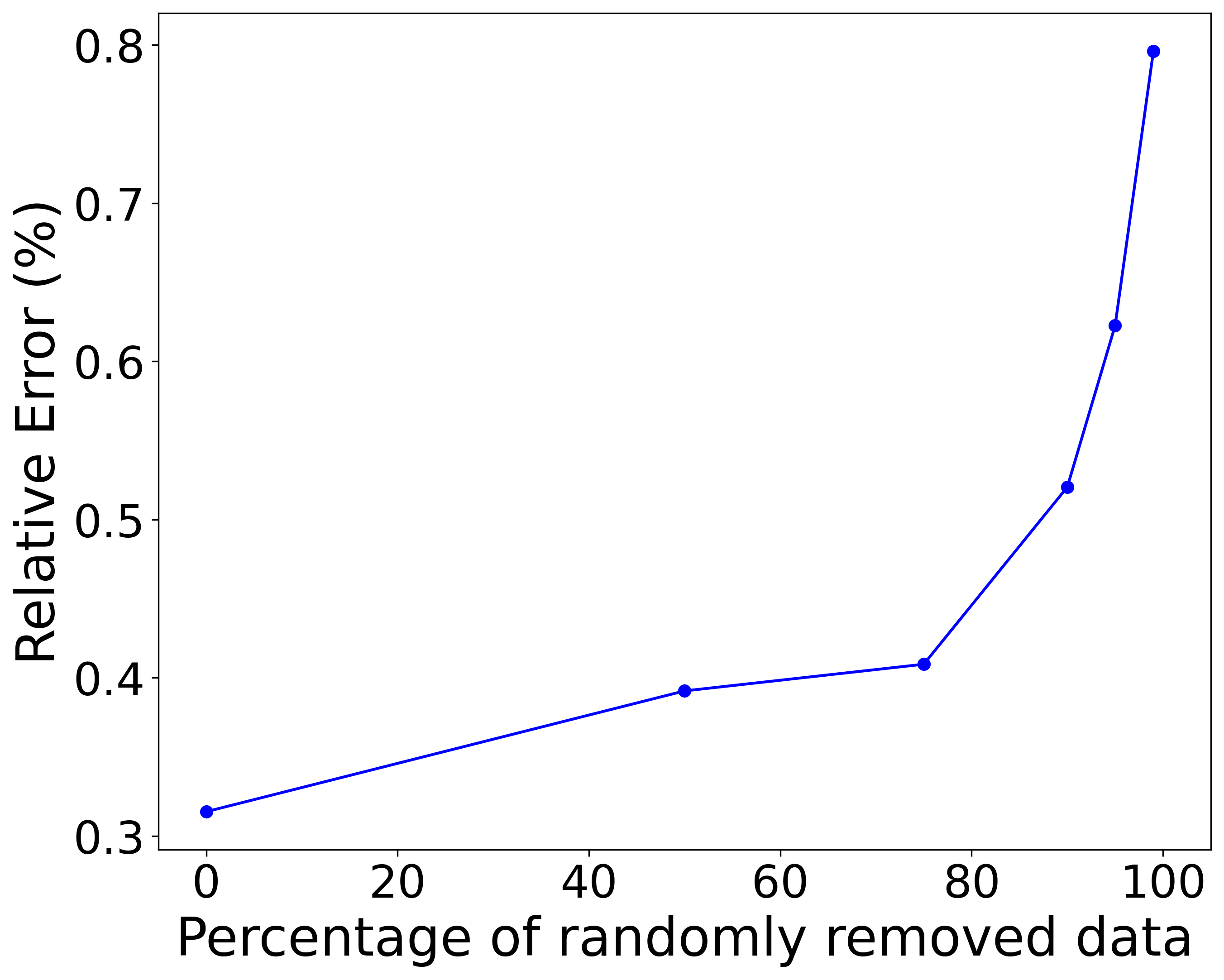}
        \caption{\centering}
        \label{fig:bx_sparse}
    \end{subfigure}
\hspace{10pt}
    \begin{subfigure}[b]{0.53\textwidth}
\centering 
        \includegraphics[width=\textwidth]{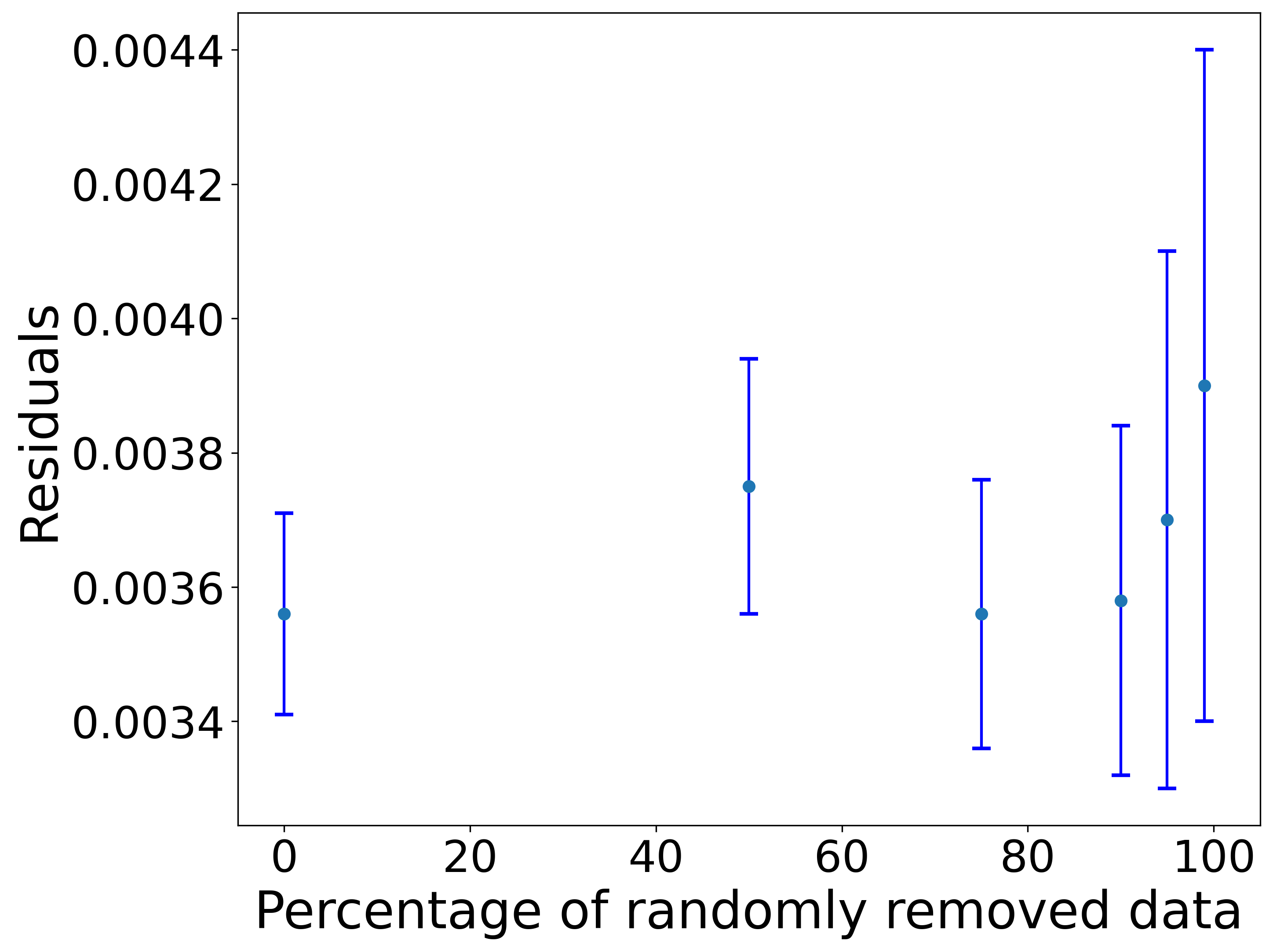}
        \caption{\centering}
        \label{fig:bx_res}
    \end{subfigure}
    \vspace{-9pt}
    \caption{
(\textbf{a}) Average of the relative error in the calculation of the x-component of the MF as a function of the sparsity of the~data. (\textbf{b}) The residuals (see Equation \eqref{resd} in \ref{application})  and their statistical error of the calculation of the left plot as a function of the sparsity of the~data. The~data used in these plots are presented in Table~\ref{tab:prior1} of \ref{table_app}. }
    \label{fig:sparse1a}
\end{figure}

Figure~\ref{fig:sparse1a} delves into the quantitative analysis of our reconstruction process. Figure~\ref{fig:sparse1a}a presents the mean value and associated statistical error of the x component reconstructed at the boundary, influenced by the level of data sparsity. A~notable trend is observed where the standard deviation error escalates in tandem with the increase in data removal, underscoring the challenges of reconstruction from sparse data. Similarly, Figure~\ref{fig:sparse1a}b charts the residuals from the reconstruction, offering a clear visualization of the error's statistical spread, again emphasizing the impact of data sparsity. \textcolor{black}{The non-zero residuals observed in Figure~\ref{fig:sparse1a}b arise primarily from the statistical nature of the reconstruction, as~the results are averaged over multiple iterations of the stochastic optimization process. Additionally, numerical approximations in solving the forward problem contribute to small discrepancies, which are expected in realistic settings. The~reported residuals, presented with their standard deviation, provide a direct measure of the uncertainty in the reconstruction process, reflecting the spread of the posterior distribution. In~addition, we calculated the confidence and prediction intervals, which are presented in \ref{CIPI}. These results demonstrate narrow confidence intervals with minimal spread even at moderate data sparsity, indicating stable and reliable reconstructions under the conditions tested.}

These results confirm the robustness of our algorithm in recovering the initial value of $\theta = 10$ (the results for each sparsity case are summarized in Table~\ref{tab:prior1} of \ref{table_app}, although~it is evident that the sparsity of the data introduces greater uncertainty, as~reflected by the widening error margins. The~completion of the reconstruction is achieved by resolving the forward problem with the estimated $\theta$, culminating in a comprehensive reconstruction despite the initial data scarcity (for an example, see Figure~\ref{FigA2}a in \mbox{ \ref{rec_mf}}). \textcolor{black}{Moreover, for~this case, we performed a sensitivity analysis proving that our procedure is robust in small changes in the previous (see \ref{application} for details). Reported residuals, presented with their standard deviation, provide a direct measure of uncertainty in the reconstruction process, reflecting the spread of the posterior distribution. In~addition, we calculated the confidence and prediction intervals, which are presented in \ref{CIPI}. These results demonstrate narrow confidence intervals with minimal spread even at moderate data sparsity, indicating stable and reliable reconstructions under the \mbox{tested conditions.}}

\vspace{-6pt}

As we reflect on the computational aspects of our proposed algorithm, it is evident that its efficiency is intricately linked to several factors. These include the volume encompassed by the cone domain, the~granularity of the solution to the forward problem dictated by the number of finite elements within the cone, and~the choice of optimization technique employed for the minimization of the negative logarithm of the posterior distribution. In~the execution of the results presented earlier, the~algorithm's computational demand was such that each iteration of Step 4 required approximately 10 min on a single core of a contemporary PC. The~structure of the algorithm lends itself to parallel processing, allowing each iteration, as~stipulated in Step 5, to~be independently executed on separate threads, thus enhancing computational throughput. The~choice of optimization method is also a critical determinant of performance. Our observations reveal that the use of dual annealing significantly reduces the run time to around 10 min, as~opposed to the 30 min necessitated by differential evolution, while maintaining a comparable level of accuracy in the results. This indicates a clear preference for dual annealing in terms of computational efficiency in this specific context. The above computations were performed on a workstation with an AMD Ryzen Threadripper PRO 5995WX Processor (manufactured by AMD), reaching 2.70 GHz to 4.50 GHz, and~equipped with 256 GB of DDR4 3200 MHz memory. The~program's computational intensity lies primarily with the CPU rather than memory, enabling efficient execution across platforms with robust multi-threading capabilities. On~a consumer-grade Intel i7 10th generation laptop, the~computational times were marginally~longer.

Furthermore, we probe the scalability of our algorithm by expanding the cone's volume twofold, keeping all other parameters constant. Each iteration of Step 4 is then extended to 30 min, suggesting a commendable scalability of our approach even as the geometry expands. This characteristic is particularly advantageous, underscoring the algorithm's potential for adaptation to larger and more complex domains without a prohibitive increase in computational requirements. More details about this can be found in \ref{complexity}.

\vspace{-4pt}


\subsection{Test Case 2: Multiple Prior~Data}

\begin{figure}[H]
\centering
    \includegraphics[scale=0.4]{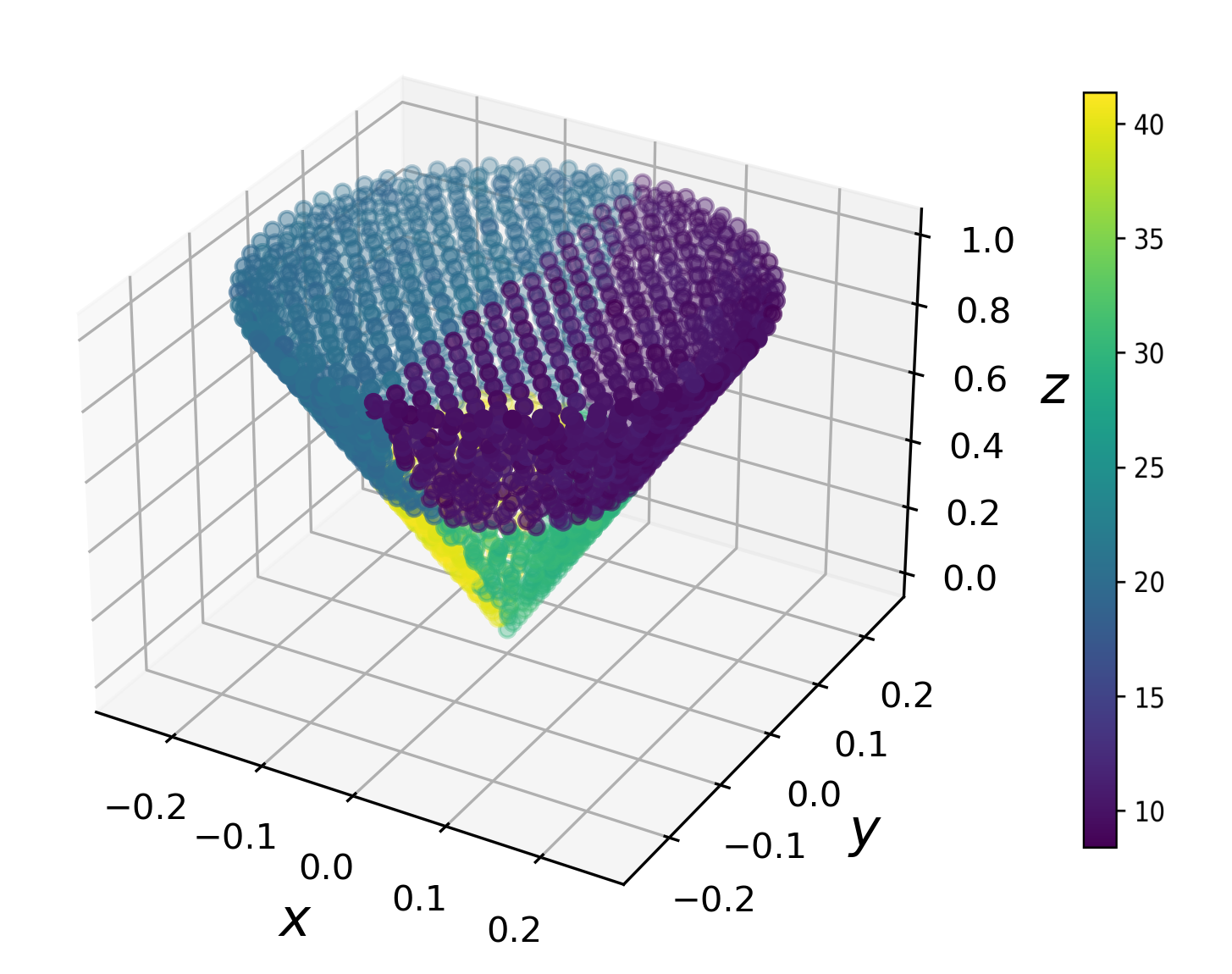}
    \caption{The  BCs for the x-component in the case of four planes to be used in the construction of synthetic data of the x-component of the MF to be used in the data-driven reconstruction calculation. They were generated from four normal distributions with different mean values and \mbox{standard deviations.}
    }
    \label{fig:4cut}
  \end{figure}

\begin{figure}[H]
\centering 
 \begin{subfigure}[b]{0.5\textwidth}
        \includegraphics[width=\textwidth]{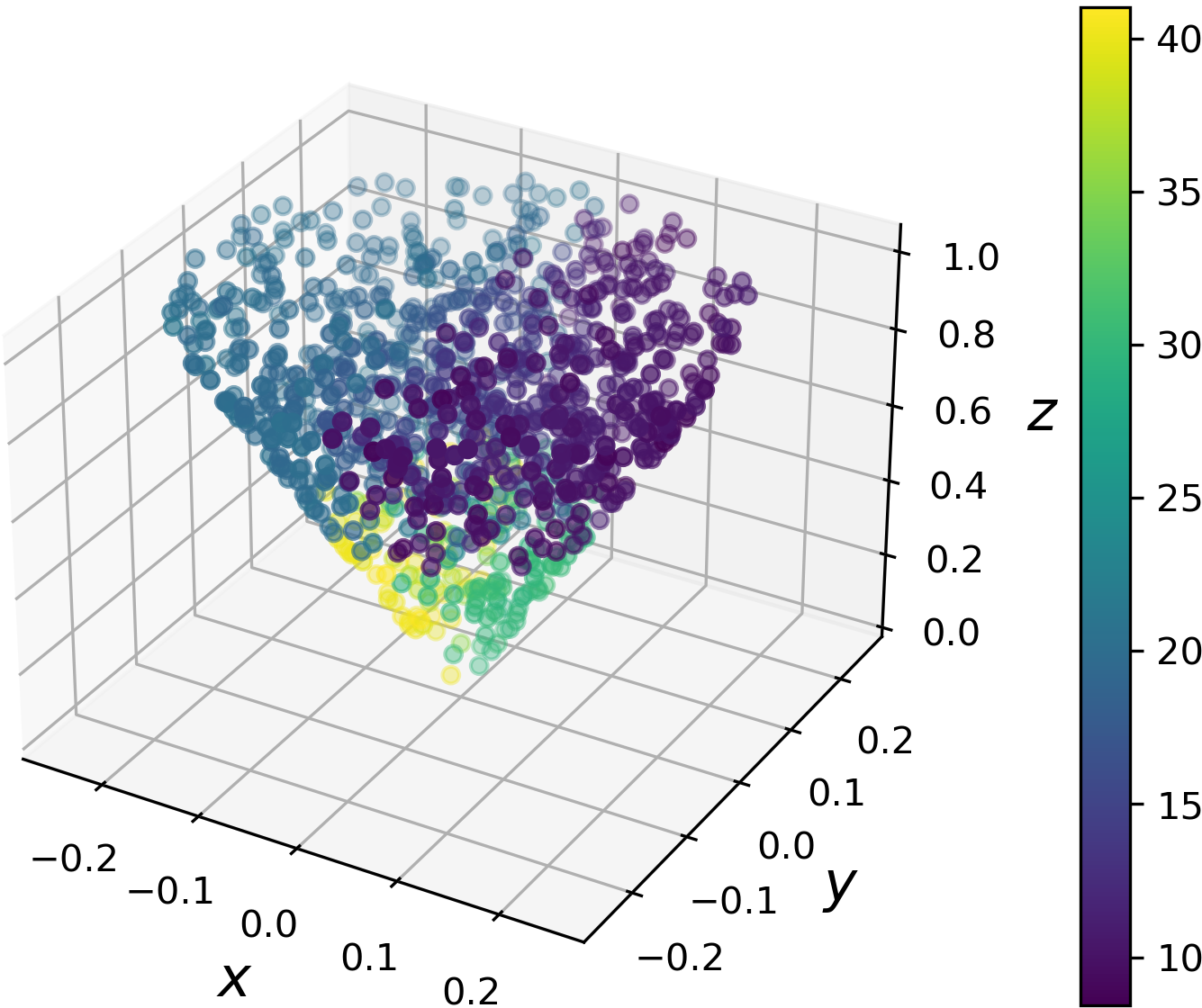}
        \caption{\centering}
        \label{fig:image1c}
    \end{subfigure}
    \hspace{25pt}
    \begin{subfigure}[b]{0.41\textwidth}
        \includegraphics[width=\textwidth]{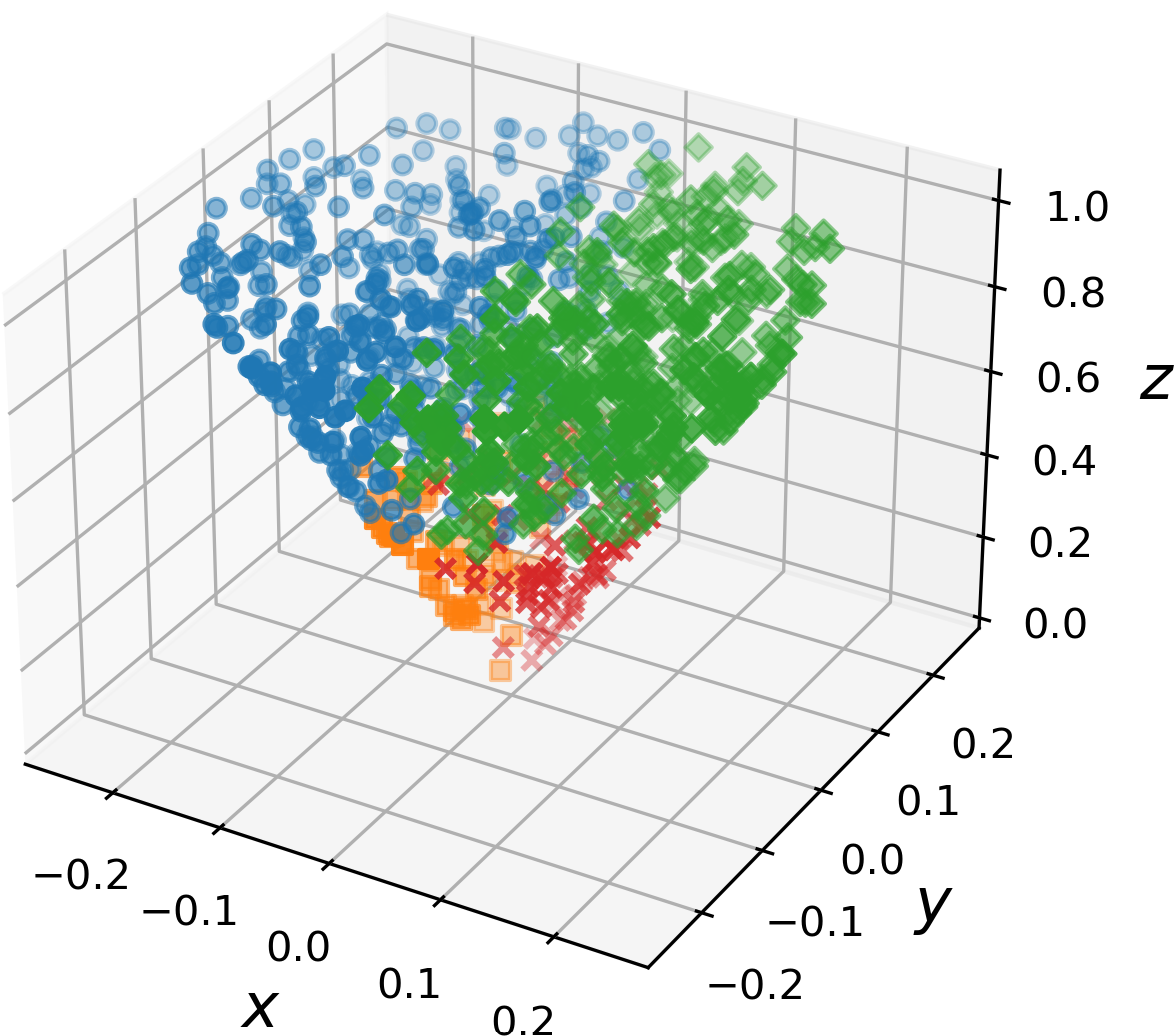}
        \caption{\centering}
        \label{fig:image2c}
    \end{subfigure}
    \vskip\baselineskip
    \begin{subfigure}[b]{0.5\textwidth}
        \includegraphics[width=\textwidth]{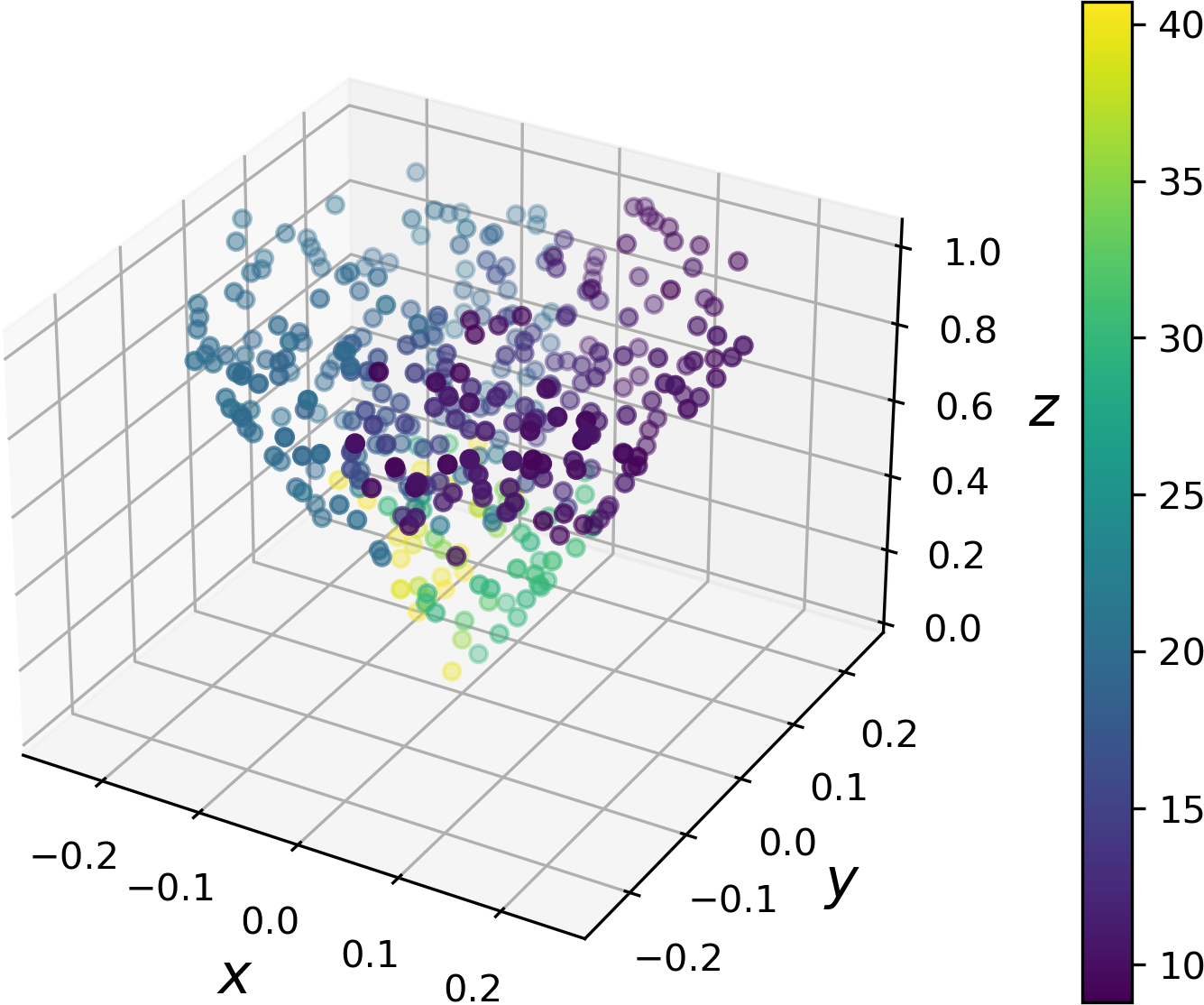}
        \caption{\centering}
        \label{fig:image3c}
    \end{subfigure}
    \hspace{25pt}
    \begin{subfigure}[b]{0.41\textwidth}
        \includegraphics[width=\textwidth]{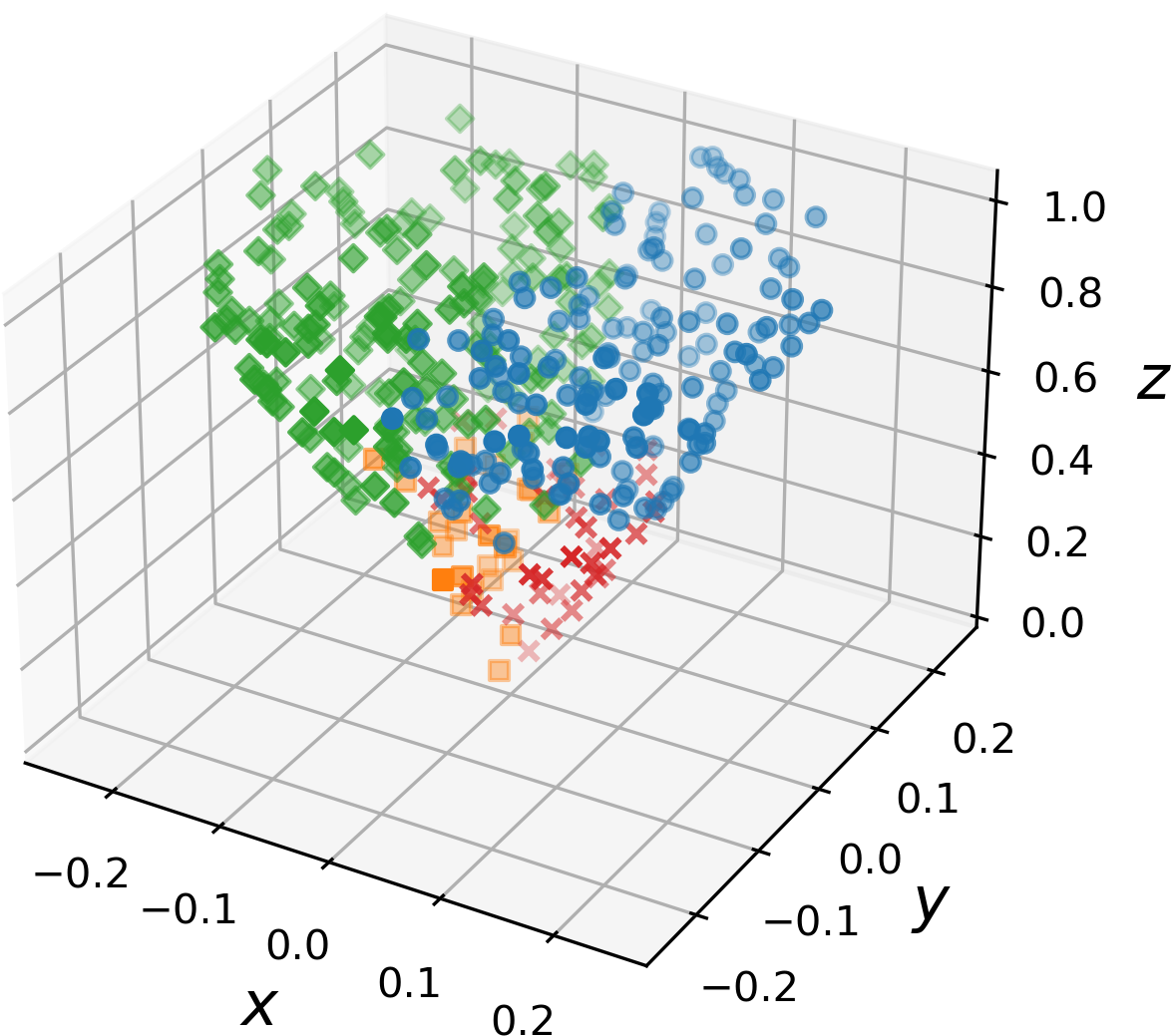}
        \caption{\centering}
        \label{fig:image4c}
    \end{subfigure}\\
            \vspace{0.5cm}
    \begin{subfigure}[b]{0.5\textwidth}
        \includegraphics[width=\textwidth]{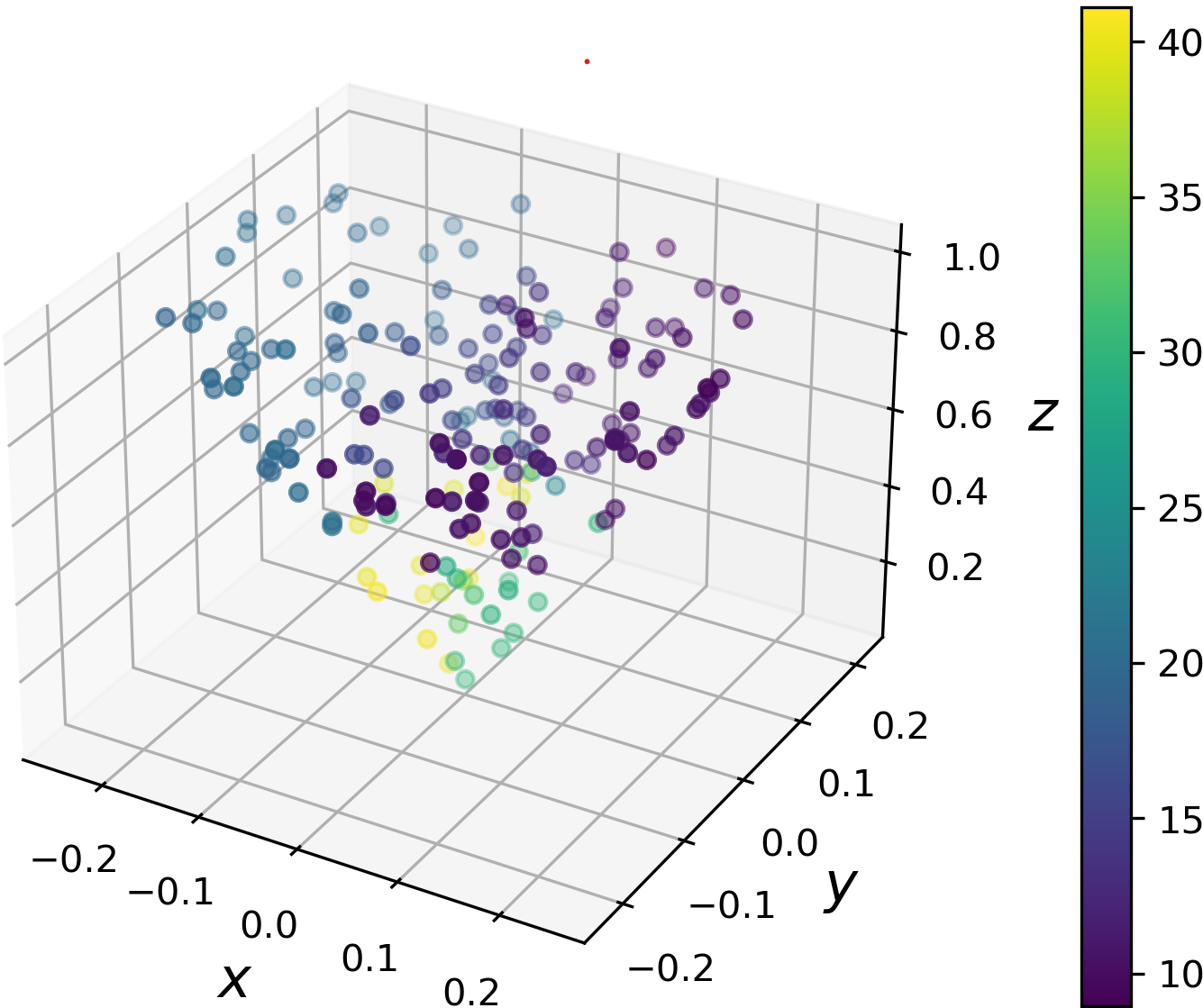}
        \caption{\centering}
        \label{fig:image5c}
    \end{subfigure}
    \hspace{25pt}
    \begin{subfigure}[b]{0.41\textwidth}
 \includegraphics[width=\textwidth]{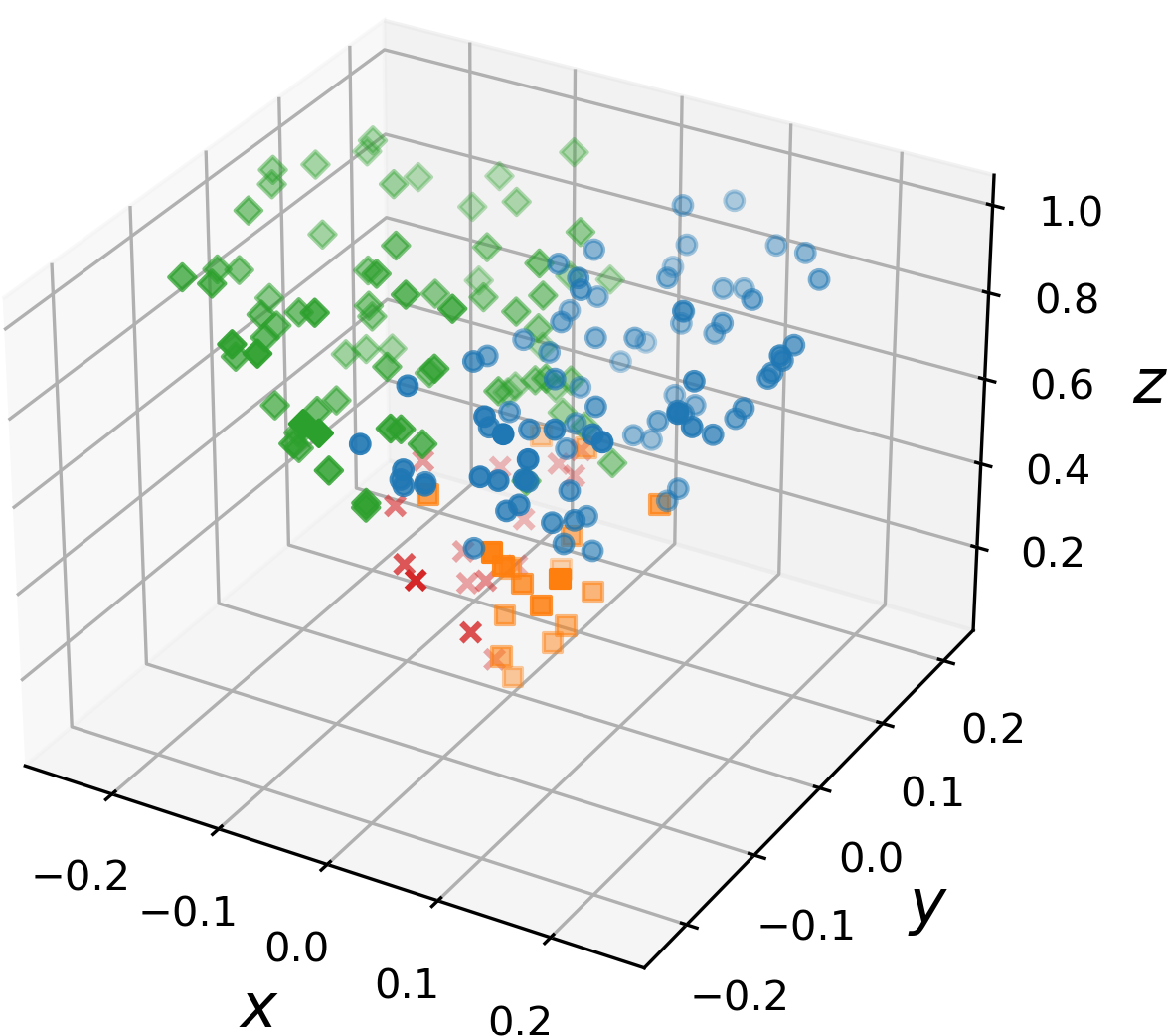}
        \caption{\centering}
        \label{fig:image6c}
    \end{subfigure}  
    \vspace{0.5cm}
    
    \caption{Left panels: synthetic data of the x-component of the MF to be used in the data-driven reconstruction calculation, as calculated from the unconstrained case-forward model with BCs from four different normal distributions (see main text). In~each subfigure, a~different percentage of data has been removed randomly to simulate a case of ``reconstruction from sparse data''. \mbox{Right panels: the} corresponding clusters used in the first step of the~algorithm. (\textbf{a}) $75\%$ of the data~removed. \mbox{(\textbf{b}) Clusters} corresponding to the left~plot. (\textbf{c}) $90\%$ of the data~removed. (\textbf{d}) Clusters corresponding to the left~plot. (\textbf{e}) $95\%$ of the data~removed. (\textbf{f}) Clusters corresponding to the left~plot.}
    \label{fig:clusters}
\end{figure}

We proceed to consider the case of a problem with sparse data by randomly removing a percentage of the solution to simulate sparse data  (the BCs are plotted in Figure \ref{fig:4cut}). The~resulting data sets can be seen in the left subfigures of Figure~\ref{fig:clusters}, that is, Figures~\ref{fig:clusters}a, \ref{fig:clusters}c and \ref{fig:clusters}e, which correspond to randomly removing $75\%$, $90\%$, and $95\%$ of the initial data, respectively. As~explained above, in general, we do not know how many distributions were responsible for creating the data. We want to associate those data in clusters, and~we perform cluster analysis to find the number of clusters and place each data point in the respective cluster. As~mentioned, our procedure is described in \ref{app_cl} and accurately predicts the number of clusters, four in this example; see the right subfigures of Figure~\ref{fig:clusters}, namely Figures \ref{fig:clusters}b \ref{fig:clusters}d and \ref{fig:clusters}f. We define the regions in the boundary as follows: in the case of four clusters, we have four boundary~regions. 

Once the regions are identified, we can proceed to solving the inverse problem, as discussed~above. 

\vspace{-6pt}

So, picking up from the meticulous cluster analysis, we advance to the inferential phase, where the four established priors are assumed to follow normal distributions with~unknown mean values and, in~this example, standard deviation fixed at unity. This assumption is the basis for developing the likelihood and posterior distributions, which subsequently inform the MAP estimation. This MAP problem is once again solved with the help of dual annealing, as~explained in Section~\ref{app_alg_gmf}, and, as~mentioned, the~whole process is analyzed in \ref{application}. The~application of the algorithm is performed three times for the three datasets shown in the right subfigures of Figure~\ref{fig:clusters}.  The~scheme very accurately returns the values of $\boldsymbol{\theta}  = \left( {10~,20~,30~,40} \right)$ for all sparsity cases considered (the results are summarized in Table~\ref{tab:prior1} of \ref{table_app}).

\begin{figure}[H]
    \begin{subfigure}[b]{0.45\textwidth}
        \centering
        \includegraphics[width=\textwidth]{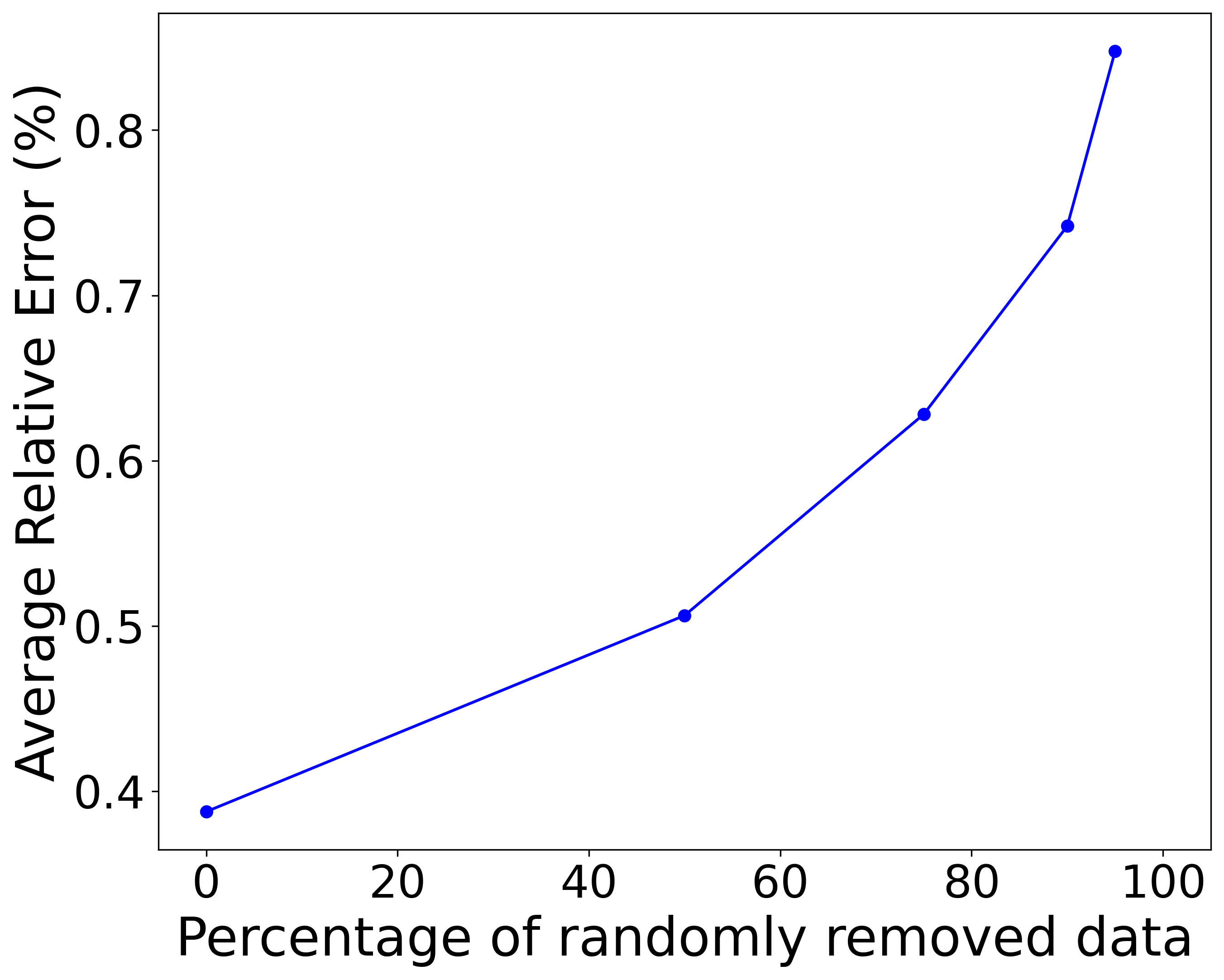}
        \caption{\centering}
        \label{fig:bx_sparse4}
    \end{subfigure}
    \hspace{10pt}
    \begin{subfigure}[b]{0.45\textwidth}
        \centering
        \includegraphics[width=\textwidth]{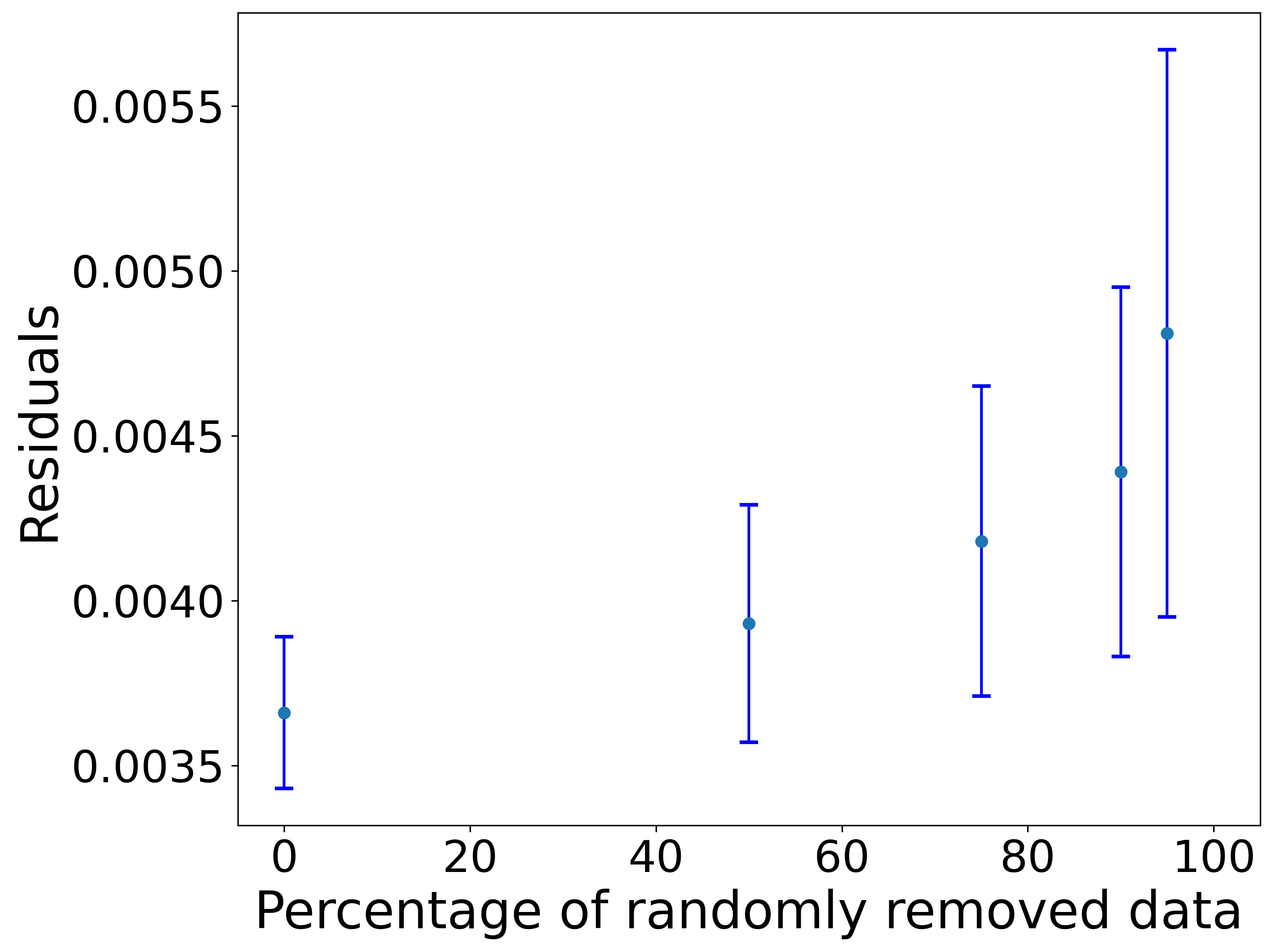}
        \caption{\centering}
        \label{fig:bx_res4}
    \end{subfigure}
    
    \vspace{0.5cm}
        
    \caption{(\textbf{a}) Average relative error in the calculation of the x-component of the MF as a function of the sparsity of the~data. (\textbf{b}) The residuals (see Equation \eqref{resd} in \ref{application}) and their statistical error of the calculation of the four plane reconstruction as a function of the sparsity of the~data. The~data used in these graphs are presented in Table~\ref{tab:prior4} of \ref{table_app}. 
    }
    \label{fig:sparse4a}
\end{figure}

Figures~\ref{fig:sparse4a}a and \ref{fig:sparse4a}b graphically represent the average of errors in each value and the residuals. The~graphical exposition reveals an expected trend: as the data becomes sparser, the~statistical error in the parameters $\boldsymbol{\theta}$ and the residuals increase, yet the initial $\boldsymbol{\theta}$ values remain robustly retrieved. The~completion of the reconstruction is achieved by resolving the forward problem with the estimated $\boldsymbol{\theta}$ (for an example, see Figure~\ref{FigA2}b in \ref{rec_mf}). \textcolor{black}{The residuals reported, along with their standard deviation, effectively capture the uncertainty in the reconstruction process by reflecting the spread of the posterior distribution. Additionally, the~calculated Confidence and prediction intervals, presented in \mbox{\ref{CIPI},} remain consistently narrow across all four prior cases, further supporting the reliability and robustness of the reconstruction method under varying data sparsity.}

When addressing the computational demands of the algorithm, the~narrative remains consistent with our previous discourse. In~the same workstation that we used in the previous example, a~single run of Step 4, devoid of parallel processing and employing dual annealing, unfolds over approximately 30 min. A~shift to differential evolution as the optimization strategy sees this duration elongate to roughly 80 min. However, the~results are comparably precise. This operational tempo is sustained even when the cone's height is doubled; the algorithm's runtime extends to a modest 90 min, a~testament to its scalable nature. We have judiciously allocated separate processing threads for each repetition, an~approach that underscores the algorithm's parallelizable~potential.

\vspace{-6pt}

In summary, the~algorithm we propose stands validated by its ability to reliably ascertain the boundary conditions' statistical attributes under the stringent test of sparse data. Using the values that were successfully recovered, we can calculate the reconstructed~field.

\vspace{-6pt}

\section{Conclusions and~Perspectives}
\label{con}

In this work, we developed a comprehensive data-driven Bayesian methodology capable of deducing the statistical characteristics of boundary conditions from sparse data. This approach is tailor-made for reconstruction calculations in complex systems. The~crux of our method lies in the implementation of a Bayesian framework that uses MAP estimation to converge on the most probable parameter values given the observed~data.

One of the primary strengths of our algorithm is its reliance on Bayesian inference coupled with MAP. This combination offers several advantages: integrating prior knowledge about parameters into the estimation process provides a probabilistic measure of uncertainty and helps overcome issues associated with poorly posed inverse problems. By~focusing on the MAP estimation, our approach effectively balances the influence of the prior distribution with the likelihood of the observed data, leading to more robust and credible parameter recovery, especially in cases with limited or noisy~data.

Furthermore, the~versatility of our algorithm is evidenced by its compatibility with any PDF. This flexibility allows it to be applied to a broad spectrum of problems where the underlying PDFs of the boundary conditions may not be normally distributed. The~ability to work with any PDF enhances the algorithm's applicability across various scientific fields and ensures that it can adapt to the unique distributions inherent to different \mbox{physical~phenomena.}

Central to our method is a clustering algorithm that intelligently groups sparse data points. This strategic grouping is essential for estimating the number of normal distributions that build the boundary conditions. Each cluster informs a prior distribution precisely defined by the cluster's mean and standard deviation. Such a setup enables a sophisticated parameter estimation through the use of stochastic optimization methods, notably dual annealing, which is particularly beneficial in~situations where data availability is a~luxury.

Delving into the results of our simulation studies, the~efficacy of the algorithm becomes clear. We successfully applied it to reconstruct the MF within a conical domain using synthetic data with a variable percentage of sparsity. Our results were encouraging, demonstrating the algorithm's ability to recover the mean values of the distributions defining the boundary conditions with high accuracy. These findings underscore the potential of our methodology to accurately infer unknown parameters and reconstruct the physical quantities of interest, even when the available data are sparse or unevenly distributed. In~this work, our calculations, while centered on normal distributions for the x-component of MF boundary conditions, represent only a fraction of possible scenarios. The~real test lies in expanding this method to include a broader spectrum of boundary conditions and evaluating its efficacy in real-world~situations.

Looking ahead, we aim to extend the application of our algorithm to an important astrophysical challenge: the reconstruction of the Milky Way magnetic field~\cite{Tsouros2023, Tsouros2024}. Using sparse data from magnetohydrodynamic simulations and direct observational input~\cite{Hutschenreuter2022}, we will evaluate the quality of reconstructions offered by our method. \textcolor{black}{This is planned for future work when abundant observational data become available.} A particularly intriguing aspect of this future work will be the backtracking of ultra-high-energy cosmic rays~\cite{Magkos2019} to their extragalactic origins, made possible by the accurate mapping of the galactic MF. In~addition to this application, a~particularly intriguing aspect of future work will be the exploration of Bayesian machine learning methods, as~hinted at in recent works like~\cite{Christofi2024}. Comparisons with methods that use MCMC algorithms will be discussed in a forthcoming~study.

In summary, the~algorithm presented in this paper not only showcases the effective use of Bayesian inference with MAP in sparse data environments but also promises a wide range of applications in disciplines where understanding the nuances of boundary conditions is paramount. The~adaptability and efficiency of our method open new avenues for research, offering potential advancements in fields far beyond the scope of this initial~study.

\section*{Acknowledgments}
G.E.P. acknowledges support from the Foundation for Research and Technology - Hellas Synergy Grant Program through the MagMASim project, jointly implemented by the Institute of Applied and Computational Mathematics and the Institute of Astrophysics.

Additionally, the algorithms and code utilized in this study will be made publicly available for the comminity on GitHub to facilitate further research and collaboration. Interested readers can access these resources at \url{https://github.com/gepavlou}.

\appendix

\section{Clustering}
\label{app_cl}
\unskip

\subsection{k-means}

We perform cluster analysis using the k-means method~\cite{Kaufman1990, Hastie2009}. The general algorithm behind the k-means method is as follows:

\begin{enumerate}
\item Choose the number of clusters (that is, two).
\item Assume two random points anywhere near the data and consider them as the center of two clusters (centroids).
\item Find the distance of every data point from both centroids.
\item Assign every data point to the centroid to which it is nearest, hence making \mbox{two clusters.}
\item Calculate the center of both the formed clusters and shift the centroids there.
\item Go to Step 3 and repeat the process until there is no change in the clusters formed.
\end{enumerate}

Obviously, it is not always clear how many clusters of data there are. The~number of clusters that we choose for a given data set cannot be random. We find their number using the silhouette~method.

\subsection{The Silhouette~Method}
\label{silhouette}

The silhouette method~\cite{Rousseeuw1987} is a method to find the optimal number of clusters and to interpret and validate consistency within clusters of data. The~silhouette Method computes silhouette coefficients for each point, which measure how much a point is similar to its own cluster compared to other clusters by providing a visual graphical representation of how well each object has been classified. The~silhouette value is a measure of how similar an object is to its own cluster (cohesion) compared to other clusters (separation). The~silhouette value ranges between $(-1, 1)$, where a high value indicates that the object is well matched to its own cluster and poorly matched to neighboring clusters. If~most objects have a high value, then the clustering configuration is appropriate. If~many points have a low or negative value, then the clustering configuration may have too many or too few~clusters.

To find the silhouette coefficient of the $i$-th~point:
\begin{enumerate}
\item Compute the average distance of that point with all other points in the same \mbox{cluster, $a(i)$.}
\item Calculate the average distance of that point with all points in the group closest to its group, $b(i)$.
\item The silhouette coefficient is defined as follows:
\begin{equation}
s(i) = \frac{b(i) - a(i)}{\max(b(i), a(i))}.
\end{equation}
\item After computing the silhouette coefficient for each point, average it out to get the silhouette score.
\end{enumerate}

We repeat the calculation for a given possible number of clusters, that is, \mbox{from~2 to 10,} for~example. The~optimal number of clusters is the one that maximizes the average silhouette coefficient across all data~points.

\section{Optimization~Methods}
\label{app3}

In this appendix, we provide a brief overview of two optimization methods used in our study: dual annealing and differential evolution. These methods offer powerful tools for global optimization in various~contexts.

Differential evolution~\cite{Lampinen} is a global optimization algorithm inspired by genetic algorithms. The~purpose of this method is to discover the global minimum of a function within a bounded search space. The~algorithm maintains a population of candidate solutions (vectors) within the specified bounds and iteratively evolves this population through mutation, crossover, and~selection operations to generate new candidate solutions for the next generation. The~distinctive features of differential evolution include mutation, crossover, and~selection. Mutation perturbs the candidate solutions by adding a differential vector, which is computed from the difference between randomly chosen individuals in the population. This mechanism promotes exploration of the search space. The~crossover combines information from multiple individuals to generate offspring, and~the specific crossover operator determines the exchange of information between individuals. Finally, selection identifies the best individuals from the current generation and the newly generated offspring to form the next generation, thereby enhancing convergence towards the global minimum. Differential evolution is highly effective in finding the global minimum of continuous and multimodal functions within a bounded search space. It is particularly suited for scenarios where the objective function is smooth and exploration of a wide range of solutions is~necessary.

\section{Minimization of the Constrained~Action}
\label{motcl}

Here, we will show the minimization of the constrained action:
\begin{equation}
\label{lagrasm}
J(\mathbf{B},\lambda )=\frac{1}{2}\int_{\Omega}{d\mathbf{x}}\left| \nabla \mathbf{B} \right|^2-\int_{\Omega}{d\mathbf{x}}\boldsymbol{\rho }\cdot \mathbf{B}+\int_{\Omega}{d\mathbf{x}\lambda \nabla \cdot \mathbf{B}},
\end{equation}
where $\lambda$ is a Lagrange multiplier one can use to find the following differential equations:
\begin{equation}
\label{eq1sm}
-\nabla ^2\mathbf{B}(\mathbf{x})=\nabla \lambda +\boldsymbol{\rho }(\mathbf{x}),~~\mathbf{x}\in \Omega, 
\end{equation}
and
\begin{equation}
\label{eq2sm}
\nabla  \cdot {\mathbf{B}}({\bf{x}}) = 0. 
\end{equation}

Using calculus of variations, we have the following:
\begin{equation}
\begin{split}
J(\mathbf{B}+\delta \mathbf{B},\lambda +\delta \lambda )&=\frac{1}{2}\int_{\Omega}{d\mathbf{x}}\left| \nabla \left( \mathbf{B}+\delta \mathbf{B} \right) \right|^2-\int_{\Omega}{d\mathbf{x}\boldsymbol{\rho }\cdot \left( \mathbf{B}+\delta \mathbf{B} \right)} \\ &
+\int_{\Omega}{d\mathbf{x}\left( \lambda +\delta \lambda \right) \nabla \cdot \left( \mathbf{B}+\delta \mathbf{B} \right)}= \\ & 
=\frac{1}{2}\int_{\Omega}{d\mathbf{x}}\left( \left| \nabla \mathbf{B} \right|^2+2\nabla \mathbf{B}\cdot \nabla \delta \mathbf{B} \right) -\int_{\Omega}{d\mathbf{x}\boldsymbol{\rho }\cdot \left( \mathbf{B}+\delta \mathbf{B} \right)}
\\ &
+\int_{\Omega}{d\mathbf{x}\left( \lambda \nabla \cdot \mathbf{B}+\delta \lambda \nabla \cdot \mathbf{B}+\lambda \nabla \cdot \delta \mathbf{B} \right)}+\mathcal{O} \delta ^{\left( 2 \right)}
\end{split}.
\end{equation}

From the above equation we can find the variation of the Lagrangian as follows:
\begin{equation}
\begin{split}
\delta J=0 & 
=J(\mathbf{B}+\delta \mathbf{B},\lambda +\delta \lambda )-J(\mathbf{B},\lambda )=
\\ &
=\int_{\Omega}{d\mathbf{x}}\nabla \mathbf{B}\cdot \nabla \delta \mathbf{B}-\int_{\Omega}{d\mathbf{x}\boldsymbol{\rho }\cdot \delta \mathbf{B}}+\int_{\Omega}{d\mathbf{x}\lambda \nabla \cdot \delta \mathbf{B}}+\int_{\Omega}{d\mathbf{x}\delta \lambda \nabla \cdot \mathbf{B}}=
\\ &
=-\int_{\Omega}{d\mathbf{x}}\left( \nabla ^2\mathbf{B}+\boldsymbol{\rho }+\nabla \lambda \right) \cdot \delta \mathbf{B}+\int_{\Omega}{d\mathbf{x}\left( \nabla \cdot \mathbf{B} \right) \delta \lambda}=0
\end{split},
\end{equation}
or
\begin{equation}
\begin{array}{l}
-\nabla ^2\mathbf{B}=\boldsymbol{\rho } +\nabla \lambda ~\mathrm{and}~\nabla \cdot \mathbf{B}=0
\end{array}.
\end{equation}

\section{Details in the Implementation of the Algorithm for the Reconstruction of the Magnetic~Field}
\label{application}

By following our algorithm, we have the following steps:
Input: Synthetic dataset $\mathbf{y}$ (sparse or not) simulating measurements or observations of the x-component of the magnetic field (MF), the~3D Laplace partial differential equation accompanied with boundary conditions of the Dirichlet type, where those of the x-component depend on the unknown parameters $\boldsymbol{\theta}$, and a clustering algorithm. 
The algorithm is as~follows:
\begin{enumerate}
\item In the general case, we do not know the true number of normal distributions that make up the boundary conditions. This is true in our example if the synthetic data are created after dividing the boundary of the cone into planes. To~address this, we use a clustering approach to group the sparse data points based on their proximity and value to each other (see  \ref{app_cl}). The~number of clusters $n_c$ then corresponds to the number of normal distributions that make up the boundary conditions. We associate each cluster with a prior distribution with each point in the boundary associated with the cluster spatially closer to it. The~unknown parameters $\boldsymbol{\theta}$ are the mean values of the aforementioned normal distributions. In~this work, we assume that the means of the normal distributions in the boundary conditions are independent. We choose as their values the mean of the data of each respective cluster, $\bar{y}_i$, while for standard deviations, we consider them all equal to one. So, the~total prior is defined as a product:
\begin{equation}
p\left( \boldsymbol{\theta} \right) =\prod_{i=1}^{n_c}{p\left( \theta _i \right)}=\prod_{i=1}^{n_c}{\frac{1}{\sqrt{2\pi}}e^{-\frac{\left( \theta _i-\bar{y}_i \right) ^2}{2}}},
\end{equation}
with $\theta$ as our unknown parameters, representing the mean value of the x-component of the MF in each~plane.

\item Consider then a initial guess for $\boldsymbol{\theta}$. By~solving the forward problem we define the cost function as follows:
\begin{equation}
J\left( \boldsymbol{\theta} \right) =\frac{1}{n_y}\sum_{i=1}^{n_y}{(y_i-f(\mathbf{x} ; \boldsymbol{\theta} ))}.
\end{equation}

The likelihood is considered to be Gaussian:
\begin{equation}
y_i-f(\mathbf{x} ; \boldsymbol{\theta} ) \sim {\cal N}(0,1),~~~i=1,...,n_y,
\end{equation}
or
\begin{equation}
p(\mathbf{y}|\boldsymbol{\theta} )=\prod_{i=1}^{n_y}{\frac{1}{\sqrt{2\pi}}}\exp \left( -\frac{(y_i-f(\mathbf{x} ; \boldsymbol{\theta} ))^2}{2} \right).
\end{equation}

\item The logarithm of the posterior is as follows:
\begin{equation}
\ln p\left( \boldsymbol{\theta} |\mathbf{y} \right) =-\sum_{j=1}^{n_y}{\frac{\left( y_i-f(\mathbf{x};\boldsymbol{\theta} ) \right) ^2}{2}}-\sum_{j=1}^{n_{\boldsymbol{c}}}{\frac{\left( \theta _i-\bar{y}_i \right) ^2}{2}}+\mathrm{constants}.
\end{equation}

\item The maximum a posteriori problem is defined as shown in the main text. So, to~find the maximum a posteriori estimate of the model parameters, we minimize the negative logarithm of the posterior distribution using stochastic optimization methods such as dual~annealing.

After the dual annealing minimization method converges to a global minimum and we find a first approximation for $\boldsymbol{\theta}$, we can also calculate the residuals:
\begin{equation}
\label{resd}
R = \frac{1}{{{n_y}}} \sum\limits_{i = 1}^{{n_y}} {\int\limits_\Omega  {d{\bf{x}}} {{\left( {{y_i} - f(x;{\boldsymbol{\theta}})} \right)}^2}}.
\end{equation}

After that, we repeat the calculation and average the results, thus finding $ \boldsymbol{\bar \theta}$ and $\bar R$ and their respective statistical errors. Provided the calculation is successful, we should recover the mean value of the parameters that were used to generate the synthetic data, i.e.,~the results of the optimization are the inferred mean values of the normal distributions that make up the boundary~conditions.

Our proposed inverse problem algorithm should then recover the mean values of these~distributions. 

\item A solution of the forward model gives the MF in the whole cone domain, thus completing the reconstruction~calculation.

\end{enumerate}

Therefore, we have as output the parameters $\boldsymbol{\bar \theta}$ and the reconstruction of the MF in the whole~domain.

\textcolor{black}{We performed a sensitivity analysis by varying the standard deviation ($\sigma$) of the normal prior while keeping the mean fixed. For~this check, we chose the case of when 50\% of the data were removed to generate the test dataset. The~reconstructed mean values and residuals remained consistent across a range of $\sigma$ values ($1.1$ to $3.0$), indicating that the results are robust to changes in the prior strength. This stability suggests that the choice of the prior standard deviation does not overly influence the reconstructed parameters, confirming the suitability of the normal prior for this inverse problem.}

\section{Tables with the Results that Were Omitted in the Main~Text}
\label{table_app}

\begin{table}[H]
\centering
\begin{tabular}{|c|c|c|}
\hline
Percentage of data we remove & $\bar\theta$         & Residuals          \\ \hline
$0\%$                            & $10.00342\pm0.03155$   & $0.00356\pm0.00015$  \\ \hline
$50\%$                          & $9.98887\pm 0.03913 $   & $0.00375\pm0.00019$  \\ \hline
$75\%$                          & $10.01981\pm0.04094$   & $0.00356\pm0.00020$  \\ \hline
$90\%$                          & $9.99900\pm0.05205$     & $0.00358\pm0.00026$  \\ \hline
$95\% $                        & $9.99473\pm0.06223$    & $0.00370\pm 0.00040$ \\ \hline
$99\%$                          & $9.98432\pm0.07948$    & $0.00390\pm0.00050$  \\ \hline
\end{tabular}
\caption{Results of the algorithm in the case of one prior. The initial value $\theta=10$ is successfully recovered. All results shown are accompanied with the respective statistical errors.}
\label{tab:prior1}
\end{table}
\vspace{-6pt}

\begin{table}[H]
\centering
\begin{tabular}{|c|l|c|}
\hline
Percentage of data we remove & \multicolumn{1}{c|}{$\boldsymbol{\bar\theta}$}         & Residuals                          \\ \hline
\multirow{4}{*}{$0\%$}          & \multicolumn{1}{c|}{$10.00537\pm0.05346$} & \multirow{4}{*}{$0.00366\pm0.00023$} \\ \cline{2-2}
                            & $19.98861\pm0.05561$                      &                                    \\ \cline{2-2}
                            & $29.84011\pm0.12618$                      &                                    \\ \cline{2-2}
                            & $40.02537\pm0.12592$                      &                                    \\ \hline
\multirow{4}{*}{$50\%$}         & \multicolumn{1}{c|}{$10.00061\pm0.07580$} & \multirow{4}{*}{$0.00393\pm0.00036$} \\ \cline{2-2}
                            & $19.96684\pm0.05993$                      &                                    \\ \cline{2-2}
                            & $30.11633\pm0.16152$                      &                                    \\ \cline{2-2}
                            & $39.59737\pm0.17049$                      &                                    \\ \hline
\multirow{4}{*}{$75\%$}         & \multicolumn{1}{c|}{$10.00928\pm0.08977$} & \multirow{4}{*}{$0.00418\pm0.00047$} \\ \cline{2-2}
                            & $20.01029\pm0.09527$                      &                                    \\ \cline{2-2}
                            & $29.41021\pm0.19511$                      &                                    \\ \cline{2-2}
                            & $39.70518\pm0.18892$                      &                                    \\ \hline
\multirow{4}{*}{$90\%$}         & \multicolumn{1}{c|}{$10.02409\pm0.11375$} & \multirow{4}{*}{$0.00439\pm0.00056$} \\ \cline{2-2}
                            & $19.96101\pm0.11800$                      &                                    \\ \cline{2-2}
                            & $29.88206\pm0.22677$                      &                                    \\ \cline{2-2}
                            & $39.95013\pm0.19304$                      &                                    \\ \hline
\multirow{4}{*}{$95\%$}         & \multicolumn{1}{c|}{$10.01150\pm0.12538$} & \multirow{4}{*}{$0.00481\pm0.00086$} \\ \cline{2-2}
                            & $20.01948\pm0.11637$                      &                                    \\ \cline{2-2}
                            & $30.22242\pm0.27096$                      &                                    \\ \cline{2-2}
                            & $39.45244\pm0.26072$                      &                                    \\ \hline
\end{tabular}
\caption{Results of the algorithm in the case of four priors. Initial values are successfully recovered. All results shown are accompanied with the respective statistical errors.}
\label{tab:prior4}
\end{table}

\color{black}
\section{Quantification of Uncertainty: Confidence and Prediction Intervals Based on Data~Sparsity}
\label{CIPI}
To further assess the robustness of the reconstruction method under varying data availability, we computed both confidence intervals (CI) and prediction intervals (PI) for the reconstructed mean values as a function of data sparsity. These statistical measures provide complementary insights into the reliability of the reconstruction under different levels of missing~data.

The confidence interval (CI) quantifies the uncertainty in estimating the mean of the reconstructed parameter. Specifically, a~$(1-\alpha)\%$ confidence interval defines a range where the true mean value of the parameter is expected to lie with a probability of $(1-\alpha)\%$. For~normally distributed parameters, the~CI is calculated using the following equation:
\begin{equation}
    \text{CI} = \bar{\theta} \pm z \cdot \frac{\sigma_\theta}{\sqrt{n}},
\end{equation}
where $\bar{\theta}$ is the mean value, $\sigma_\theta$ is the standard deviation of the reconstructed parameter, $n$ is the number of observations, and~$z$ is the critical value from the Student's t-distribution for a 95\% confidence~level.

The prediction interval (PI) provides a range where new reconstructed values are expected to fall, accounting for both the spread of the data and the uncertainty in the \mbox{mean estimate:}
\begin{equation}
    \text{PI} = \bar{\theta} \pm z \cdot \sigma_\theta \cdot \sqrt{1 + \frac{1}{n}}.
\end{equation}

Table~\ref{tab:prior1_CI_PI} presents the computed CI and PI for various data sparsity levels for the case of a single prior, indicating the impact of reducing the number of observations on the uncertainty of the reconstructed mean~values.

\begin{table}[H]
\centering
\renewcommand{\arraystretch}{1.2}
\begin{tabular}{|c|c|c|c|c|}
\hline
Percentage & CI Lower & CI Upper & PI Lower & PI Upper \\ \hline
0\%        & 9.9972   & 10.0096  & 9.9937   & 10.0131  \\ \hline
50\%       & 9.9811   & 9.9966   & 9.9742   & 10.0035  \\ \hline
75\%       & 10.0120  & 10.0277  & 10.0048  & 10.0349  \\ \hline
90\%       & 9.9901   & 10.0079  & 9.9835   & 10.0145  \\ \hline
95\%       & 9.9845   & 10.0049  & 9.9774   & 10.0120  \\ \hline
99\%       & 9.9704   & 9.9983   & 9.9625   & 10.0062  \\ \hline
\end{tabular}
\caption{Confidence and Prediction Intervals for the Reconstructed Mean Values as a Function of Data Sparsity for the case of a single prior. The values of $\theta$ used are shown in Table \ref{tab:prior1}.}
\label{tab:prior1_CI_PI}
\end{table}

The results presented in Table~\ref{tab:prior1_CI_PI} illustrate the effect of increasing data sparsity on the uncertainty of the reconstructed mean values. As~the data sparsity increases, both the CI and PI widen, indicating higher uncertainty in the reconstructed values. For~example, with~0\% sparsity, the~CI is narrow, suggesting a highly constrained estimate of the mean value. However, as~data sparsity reaches 99\%, both the CI and PI expand, reflecting the increased uncertainty due to reduced data availability. Despite this increase, the~intervals remain relatively constrained, suggesting the algorithm's robustness even under significant~sparsity.

The prediction intervals are consistently wider than the confidence intervals, as~expected since they account for both the variability in the data and the uncertainty in the mean estimate. These results emphasize the importance of data availability in inverse problems and demonstrate the robustness of the proposed algorithm under moderate sparsity levels while highlighting growing uncertainty with extreme data~reductions.

\begin{table}[H]
\centering
\renewcommand{\arraystretch}{1.2}
\begin{tabular}{|c|cc|cc|cc|cc|}
\hline
\multirow{2}{*}{\begin{tabular}[c]{@{}c@{}}Percentage of \\ data we remove\end{tabular}} & \multicolumn{2}{c|}{CI Prior 1}      & \multicolumn{2}{c|}{PI Prior 1}      & \multicolumn{2}{c|}{CI Prior 2}      & \multicolumn{2}{c|}{PI Prior 2}      \\ \cline{2-9} 
                                                                                         & \multicolumn{1}{c|}{Lower}  & Upper  & \multicolumn{1}{c|}{Lower}  & Upper  & \multicolumn{1}{c|}{Lower}  & Upper  & \multicolumn{1}{c|}{Lower}  & Upper  \\ \hline
0\%                                                                                      & \multicolumn{1}{c|}{9.997}  & 10.009 & \multicolumn{1}{c|}{9.993}  & 10.013 & \multicolumn{1}{c|}{19.982} & 19.995 & \multicolumn{1}{c|}{19.978} & 19.999 \\ \hline
50\%                                                                                     & \multicolumn{1}{c|}{9.981}  & 9.996  & \multicolumn{1}{c|}{9.974}  & 10.003 & \multicolumn{1}{c|}{19.959} & 19.974 & \multicolumn{1}{c|}{19.952} & 19.981 \\ \hline
75\%                                                                                     & \multicolumn{1}{c|}{10.012} & 10.027 & \multicolumn{1}{c|}{10.005} & 10.034 & \multicolumn{1}{c|}{20.001} & 20.017 & \multicolumn{1}{c|}{19.994} & 20.024 \\ \hline
90\%                                                                                     & \multicolumn{1}{c|}{9.990}  & 10.007 & \multicolumn{1}{c|}{9.983}  & 10.014 & \multicolumn{1}{c|}{19.952} & 19.969 & \multicolumn{1}{c|}{19.945} & 19.976 \\ \hline
95\%                                                                                     & \multicolumn{1}{c|}{9.984}  & 10.004 & \multicolumn{1}{c|}{9.977}  & 10.012 & \multicolumn{1}{c|}{20.012} & 20.027 & \multicolumn{1}{c|}{20.005} & 20.034 \\ \hline
\multirow{2}{*}{\begin{tabular}[c]{@{}c@{}}Percentage of \\ data we remove\end{tabular}} & \multicolumn{2}{c|}{CI Prior 3}      & \multicolumn{2}{c|}{PI Prior 3}      & \multicolumn{2}{c|}{CI Prior 4}      & \multicolumn{2}{c|}{PI Prior 4}      \\ \cline{2-9} 
                                                                                         & \multicolumn{1}{c|}{Lower}  & Upper  & \multicolumn{1}{c|}{Lower}  & Upper  & \multicolumn{1}{c|}{Lower}  & Upper  & \multicolumn{1}{c|}{Lower}  & Upper  \\ \hline
0\%                                                                                      & \multicolumn{1}{c|}{29.820} & 29.859 & \multicolumn{1}{c|}{29.807} & 29.872 & \multicolumn{1}{c|}{40.003} & 40.047 & \multicolumn{1}{c|}{39.990} & 40.060 \\ \hline
50\%                                                                                     & \multicolumn{1}{c|}{30.092} & 30.140 & \multicolumn{1}{c|}{30.075} & 30.157 & \multicolumn{1}{c|}{39.570} & 39.624 & \multicolumn{1}{c|}{39.553} & 39.641 \\ \hline
75\%                                                                                     & \multicolumn{1}{c|}{29.381} & 29.439 & \multicolumn{1}{c|}{29.364} & 29.456 & \multicolumn{1}{c|}{39.684} & 39.727 & \multicolumn{1}{c|}{39.667} & 39.744 \\ \hline
90\%                                                                                     & \multicolumn{1}{c|}{29.853} & 29.911 & \multicolumn{1}{c|}{29.836} & 29.928 & \multicolumn{1}{c|}{39.924} & 39.976 & \multicolumn{1}{c|}{39.907} & 39.993 \\ \hline
95\%                                                                                     & \multicolumn{1}{c|}{30.185} & 30.259 & \multicolumn{1}{c|}{30.161} & 30.283 & \multicolumn{1}{c|}{39.430} & 39.474 & \multicolumn{1}{c|}{39.413} & 39.491 \\ \hline
\end{tabular}
\caption{Confidence and Prediction Intervals for the Reconstructed Mean Values as a Function of Data Sparsity for the case of four priors. The values of $\theta$ used are shown in Table \ref{tab:prior4}.}
\label{tab:prior4_CI_PI}
\end{table}

The same trends are observed for the multiple-prior case (Table \ref{tab:prior4_CI_PI}). The~CI and PI remain relatively narrow for all sparsity levels but expand as the data sparsity increases, reflecting greater uncertainty due to limited~observations.

\section{Time and Space Complexity~Analysis}
\label{complexity}
To quantify the computational performance of the proposed algorithm, we conducted a series of controlled tests to evaluate both time complexity and space complexity. The~goal was to measure how the runtime and memory usage scale with increasing mesh resolution while keeping other parameters~constant.

For the complexity tests, we chose a relatively simple case. They were performed using the following~setup:
\begin{itemize}
    \item Domain: Unit cube in 3D space.
    \item Element Type: Linear tetrahedral element (\(\mathbb{P}_1\)).
    \item Boundary Conditions: Fixed using a single prior.
    \item Data Sparsity: 50\% of the synthetic data was removed.
    \item Realization Count: A single realization was performed to focus on the scaling behavior rather than statistical variation.
    \item Solver: The default direct solver (LU decomposition) from the PETSc backend in FEniCSx was used.
    \item The computations were performed on a workstation with an AMD Ryzen Threadripper PRO 5995WX Processor equipped with 256 GB of DDR4 3200 MHz memory.
\end{itemize}

\begin{figure}[H]
    \centering
    \begin{subfigure}[b]{0.45\textwidth}
        \centering
        \includegraphics[width=\textwidth]{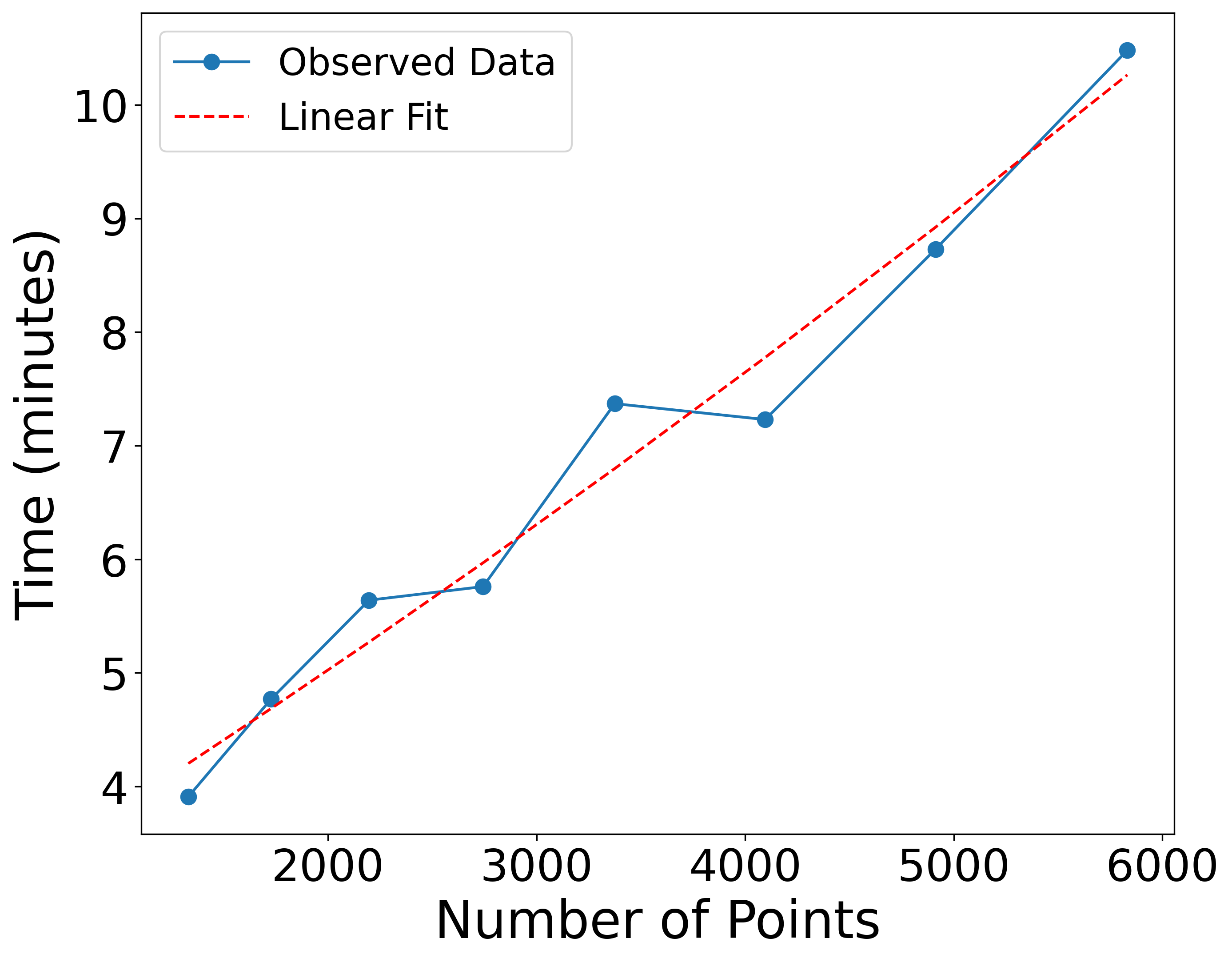}
        \caption{\centering}
    \label{fig:time_complexity}
    \end{subfigure}
    \hfill
    \begin{subfigure}[b]{0.45\textwidth}
        \centering
        \includegraphics[width=\textwidth]{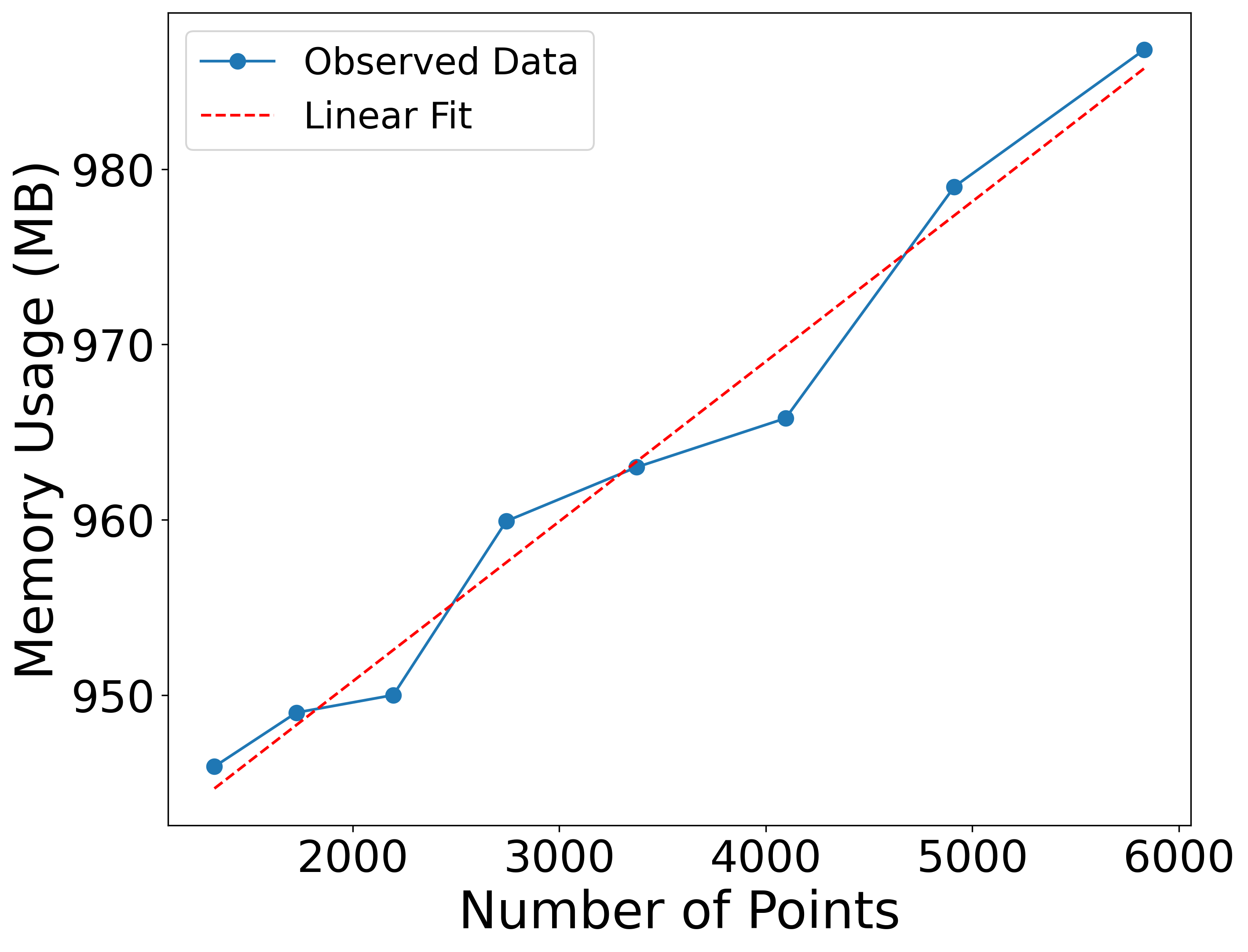}
        \caption{\centering}
    \label{fig:space_complexity}
    \end{subfigure}
    \label{fig:complexity}
\vspace{0.3 cm}
    
\caption{(\textbf{a}) Time complexity analysis: the total run time as a function of mesh~size. (\textbf{b}) Space complexity analysis: memory usage as a function of mesh~size. The~data are shown in blue dots and the red dashed line is a linear fit ($y=a+bx$) . The~linear fit closely follows the observed data points. Fit results: left plot $a = (1.344 \pm 0.098) \times 10^{-3}$, $b = 2.333 \pm 0.352$; right plot $a = (9.126 \pm 0.578) \times 10^{-3}$, $b = 932.5 \pm 2.1$. }    \label{FigA1}
\end{figure}

The runtime was measured as the total time required to solve the forward and inverse problems while varying the number of mesh elements. The~results are shown in Figure~\ref{FigA1}a.

\vspace{-6pt}

The data presented in Figure~\ref{FigA1}a indicate a behavior closer to linear scaling for the tested range of mesh resolutions. This could be attributed to the solver's optimized handling of sparse matrices for the smaller domains~tested.

The space complexity was evaluated based on the peak memory usage during the execution, as~shown in Figure~\ref{FigA1}b.

The memory usage scales linearly with the number of mesh points, which is consistent with the expected behavior for storing sparse matrices in FEM. The~small error margins on the fit parameters reinforce the reliability of this observation. However, in~practical applications like the ones presented in the main text, the~procedure must be run several times, and the results must be averaged. Obviously, the procedure can then be run in~parallel.

\section{The Reconstructed Magnetic~Field}
\label{rec_mf}

\vspace{-12pt}
\begin{figure}[H]
    \centering
    \begin{subfigure}[b]{0.45\textwidth}
        \centering
        \includegraphics[width=\textwidth]{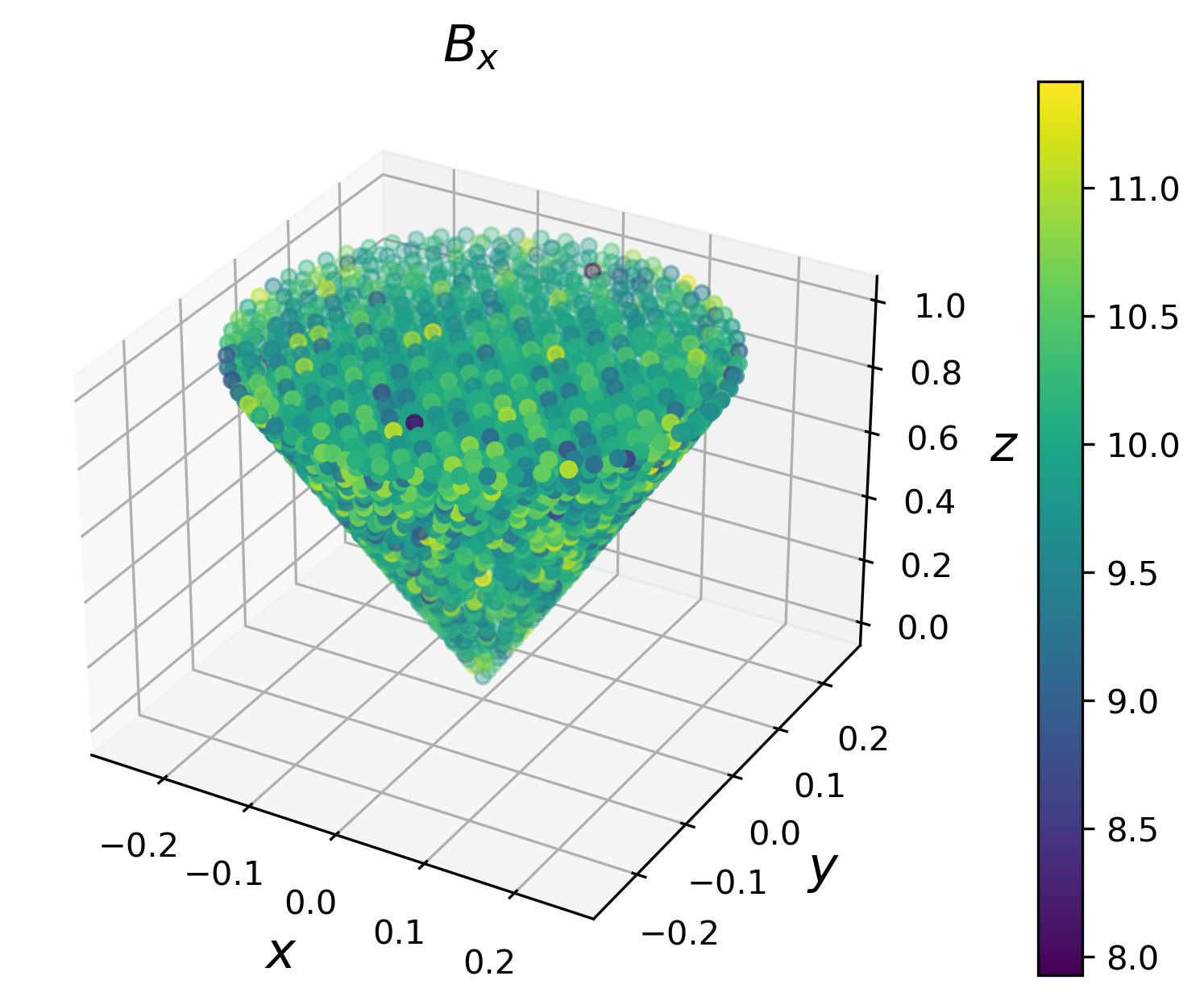}
        \caption{\centering}
        \label{fig:bxr1b}
    \end{subfigure}
    \hfill
    \begin{subfigure}[b]{0.45\textwidth}
        \centering
        \includegraphics[width=\textwidth]{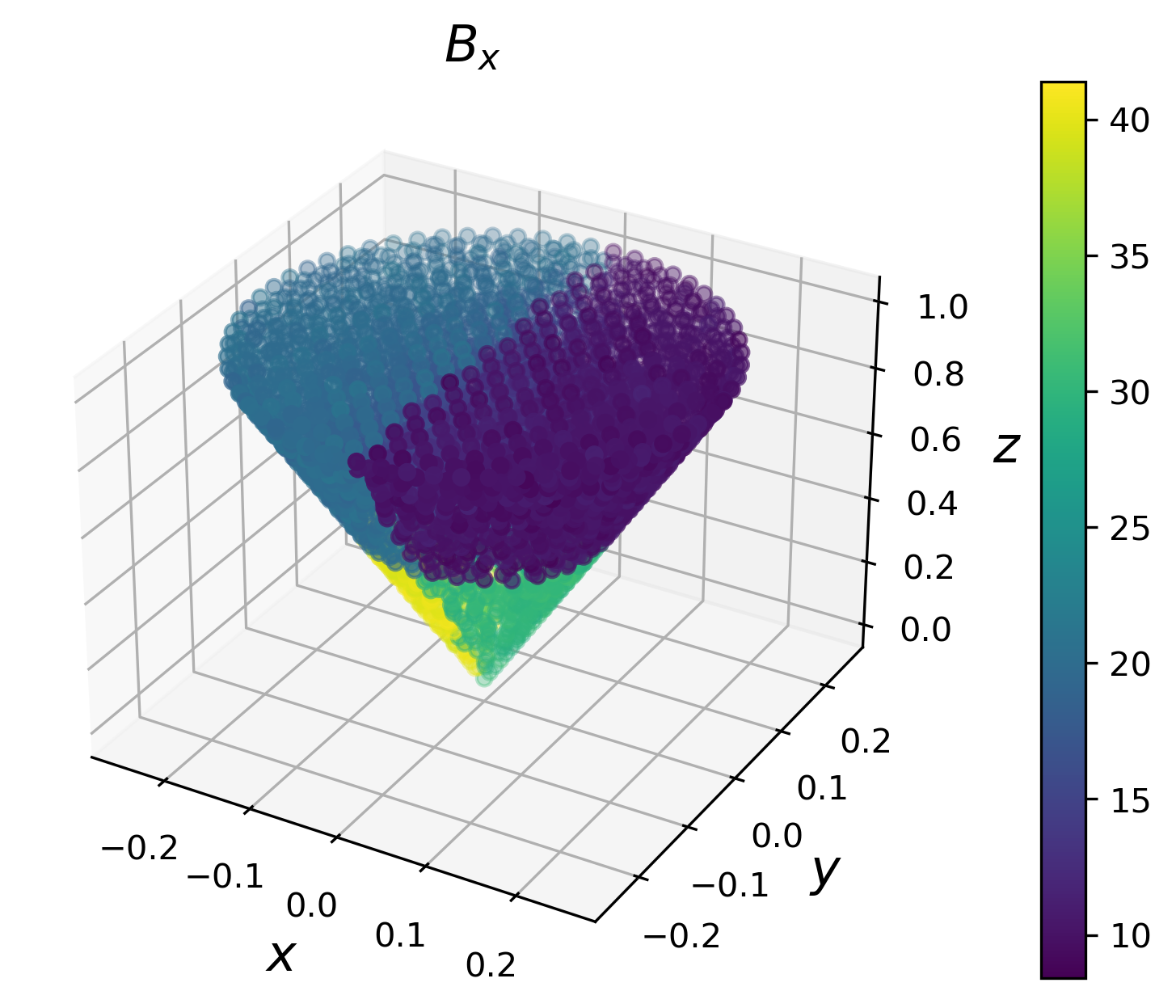}
        \caption{\centering}
        \label{fig:bxr4}
    \end{subfigure}
    \label{fig:sparse4b}

 \caption{(\textbf{a}) The reconstructed magnetic field for the case of 1 prior, where we removed 95\% of the~data. (\textbf{b}) The reconstructed magnetic field for the case of 4 priors, where we removed 95\% of the~data.} \label{FigA2}
\end{figure}

\end{document}